\documentclass[preprint,11pt]{JHEP3} 

\JHEPspecialurl{http://jhep.sissa.it/JOURNAL/JHEP3.tar.gz}

\usepackage{epsfig,multicol}

\usepackage{rotating}

\def\beq{\begin{equation}}
\def\eeq{\end{equation}}
\def\bea{\begin{eqnarray}}
\def\eea{\end{eqnarray}}

\def\eq#1{{Eq.~(\ref{#1})}}
\def\fig#1{{Fig.~\ref{#1}}}
\newcommand{\bas}{\bar{\alpha}_S}
\newcommand{\as}{\alpha_S}

\newcommand{\Lb}{\left(}
\newcommand{\Rb}{\right)}

\parskip=\itemsep               

\def\ap#1#2#3{     {\it Ann. Phys. (NY) }{\bf #1} (19#2) #3}
\def\arnps#1#2#3{  {\it Ann. Rev. Nucl. Part. Sci. }{\bf #1} (19#2) #3}
\def\npb#1#2#3{    {\it Nucl. Phys. }{\bf B#1} (19#2) #3}
\def\plb#1#2#3{    {\it Phys. Lett. }{\bf B#1} (19#2) #3}
\def\prd#1#2#3{    {\it Phys. Rev. }{\bf D#1} (19#2) #3}
\def\prep#1#2#3{   {\it Phys. Rep. }{\bf #1} (19#2) #3}
\def\prl#1#2#3{    {\it Phys. Rev. Lett. }{\bf #1} (19#2) #3}
\def\ptp#1#2#3{    {\it Prog. Theor. Phys. }{\bf #1} (19#2) #3}
\def\rmp#1#2#3{    {\it Rev. Mod. Phys. }{\bf #1} (19#2) #3}
\def\zpc#1#2#3{    {\it Z. Phys. }{\bf C#1} (19#2) #3}
\def\mpla#1#2#3{   {\it Mod. Phys. Lett. }{\bf A#1} (19#2) #3}
\def\nc#1#2#3{     {\it Nuovo Cim. }{\bf #1} (19#2) #3}
\def\yf#1#2#3{     {\it Yad. Fiz. }{\bf #1} (19#2) #3}
\def\cpc#1#2#3{    {\it Comp. Phys. Commun. }{\bf #1} (19#2) #3}

\def\thefootnote{\fnsymbol{footnote}}

\title{\Large  Solution for  the BFKL Pomeron Calculus in zero transverse dimensions}
\author{\large M. 
Kozlov\thanks{Email:
kozlov@post.tau.ac.il;}\,\,\,\,and\,\, E. Levin\thanks{Email: leving@post.tau.ac.il, 
levin@mail.desy.de;} \\
Department of Particle Physics, School of Physics and Astronomy\\ Raymond and Beverly Sackler Faculty 
of Exact Science\\  Tel Aviv University, Tel Aviv, 69978, Israel}

\abstract{ In this paper the exact analytical solution is found for the BFKL Pomeron calculus in 
 zero transverse dimensions, in which all Pomeron loops have been included. 
The comparison with 
the approximate methods of  the  solution is given, and   the kinematic regions are discussed   where they describe the
behaviour of the scattering amplitude quite well. In particular, the semi-classical approach is 
considered, which reproduces the main properties of the exact solution  at large values of 
rapidity ($Y \,\geq\,10 $). It is shown that the mean field approximation leads to a good description of the scattering
amplitude only if the amplitude at low energy is rather large. However, even in this case, it does not lead to the correct 
asymptotic behaviour of the scattering amplitude at high energies. }


\preprint{  TAUP -2823-06\\
hep-ph/0604039\\
\today}
\begin{document}

\def\thefootnote{\arabic{footnote}}
\section{Introduction}
\label{sec:Int}
It is well known that the BFKL Pomeron calculus\cite{GLR,MUQI} gives the simplest and the most 
elegant approach to the high energy amplitude in QCD. This approach is based on the BFKL Pomeron  
\cite{BFKL} and the  interactions between the BFKL Pomerons \cite{BART,BRN,NP,BLV}, which are 
taken into account in the spirit of the  Gribov Reggeon Calculus \cite{GRC}. It can be 
written in the form of  the functional integral (see Ref. \cite{BRN})  and  formulated in  
terms  of the generating functional (see Refs. \cite{MUCD,L1,L2,L3,L4}). However, in spite of the 
fact that we have learned a lot about the  properties of the high energy asymptotic behaviour at 
high 
energies, even the 
simplest case of the BFKL Pomeron in zero transverse dimension (a toy model , see Refs.
\cite{MUCD,L3,L4}) has not been solved\footnote{Some progress has been made in Refs. 
\cite{L4,REST,SHXI} and we will comment  on these efforts below.} In this paper we develop the
analytical approach to this problem and find the solution.
 We hope that this solution will help us to develop a technique for the general approach
in which the amplitude depends on the size of interacting dipoles as well as on their impact 
parameters.

The simple toy model has some advantages which makes it a good laboratory for seeking the 
solutions 
that include the Pomeron loops. First of all, the mean field approximation \cite{GLR,MUQI,MV,BK} 
for this model has a simple analytical solution \cite{L1,L2} while it is a difficult problem in the 
general case and we have only few analytical approaches \cite{LT,IIM}, therefore,  we  have to rely
very 
often (too often)   on the 
numerical 
solutions \cite{THEORVRSDATA} .  The JIMWLK approach \cite{JIMWLK}  coincides with the
Balitsky-Kovchegov \cite{BK} one in
this model making our  life simpler in the mean field approach.

The first corrections due to the Pomeron loops which were 
suggested in Ref. \cite{MUPA,IM} can be easily summed in this model \cite{KOLE} giving a guide for
the general impact of the Pomeron loops on the high energy asymptotic behaviour of the 
scattering amplitude.   
 
We are aware that we can loose the important property of the QCD approach (see Refs. \cite{IT,MSW,
IST,KOLU,HIMS})  but we believe that the analytical solution to the simple model is a necessary step
in finding more general and certainly more complicated solutions. We are certain that methods 
 developed  for solving this model,  will be useful for searching for a general solution of this 
challenging and difficult problem.

The paper is organized as follows. In the next section we discuss the general properties of the
equation that we need to solve. Being a diffusion equation this equation could be rewritten as the 
Sturm-Liouville equation which has a discrete spectrum.
 In section 3 we develop the semi-classical approach which results in finding  the analytical 
solution 
to the equation. We show that this solution approaches the so called asymptotic solution at high 
energies.  In this section we discuss the analytical solution at $u \to 1$ which gives the
transparent 
example of the main  properties of the general solution.

In section 4 we  find the general analytical solution to the problem. We state that the prolate
spheroidal wave functions  is the complete set of the eigenfunctions of the master equation and 
build the Green function as well as solution to the master equation that satisfy the initial and 
boundary conditions that has been discussed in section 2.

In conclusions we discuss our results and check the approximate approaches, developed for this kind
of equations,  against the exact 
solution.

\section{The equation and its general properties}
\label{sec:GA}
\subsection{The equation and initial and boundary conditions.}
In the toy model the equation for the generating functional\footnote{Actually for the toy model, in 
which there is no dependence on the sizes of interacting dipoles,  the generating functional 
degenerates to the generating function.}
has a simple form 
\cite{MUCD,L3,L4}
\beq \label{GA1}
\frac{\partial Z}{\partial{\cal Y}}\,\,=\,\,\,u\,(1 - u)\,\left(-\,\,\kappa\,\,
\frac{\partial 
Z}{\partial
u}\,\,+\,\,
\frac{\partial^2 Z}{\partial u^2} \right)
\eeq
 where 
\beq \label{CALY}
{\cal Y} \,=\,\,\Gamma(2 \to 1)\,Y \,=\,(\Gamma(1 \to2)/\kappa)\,Y
\eeq
 with vertex $\Gamma(2 \to 1)$ which describes 
a 
merging of two dipoles into one dipole. $\kappa$ is the large parameter of our problem and it is 
equal to
\beq \label{KAPPA}
\kappa\,\,\,=\,\,\,\frac{2 \, \cdot \, \Gamma(1 \to 2)}{\Gamma(2 \to 1)}\,\,=\,\,\frac{1}{\as^2}\,\,\,\gg\,\,1
\eeq
where $\as$ is QCD coupling and $\Gamma(1 \to 2)\,\,=\,\,(N_c/\pi)\as\,\,=\,\,\bas $ is the vertex 
for the decay of the dipole into two 
dipoles. The order of these vertices $\Gamma(1 \to 2)$ and $\Gamma(2 \to 1)$ are chosen from the 
general equation for the generating functional in QCD (see Ref. \cite{L3} for example).

\eq{GA1} is a typical diffusion equation with the diffusion coefficient that depends on variable 
$u$. We need to add an initial and boundary condition to solve such kind of equation. Generally, this 
equation describes the interaction between Pomerons with intercept $\Delta_P = \Gamma(1 \to 2) 
=\bas$. The Pomeron diagrams that describes \eq{GA1} are shown in \fig{pom}.
\FIGURE[ht]{
\centerline{\epsfig{file=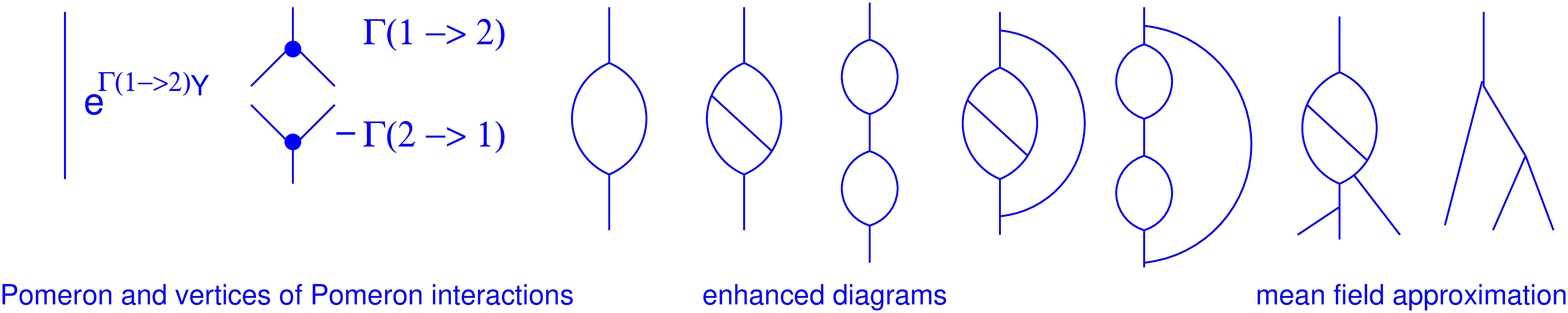,width=160mm}}
\caption{The Pomeron diagrams that are summed by  \eq{GA1}. Solid lines denote the Pomerons. The 
first three diagrams show the Pomeron and vertices of Pomeron interaction. The next five 
diagrams are the examples of so called enhanced diagrams which we are going to sum in this paper. 
The last two diagrams give the example of the Pomeron diagrams in the mean field  approximation.}
\label{pom} }

In this paper we are going to calculate the sum of enhanced diagrams (see \fig{pom}) and for them 
we have the following initial condition \footnote{In \eq{GAIN} as well everywhere below we 
use the notation $Y$ for ${\cal Y}$. Hopefully it will not lead to troubles in understanding.}
\beq \label{GAIN}
Z(Y = 0; u) \,\,=\,\,u
\eeq

Since the physical meaning of $Z$ is
\beq \label{GAZ}
Z(Y;u)\,\,=\,\,\sum_{n = 0}^{\infty}\,\,P_n(Y)\,u^n
\eeq
where $P_n(Y)$ is the  probability to find $n$-dipoles (or/and $n$-Pomerons) at rapidity $Y$.
Therefore, \eq{GAIN} means that at low energies we have only one Pomeron.

The master equation (see \eq{GA1}) describes Markov's chain for the  probabilities $P_N(Y)$, namely
\cite{IT,L3}
\bea \label{MARCH}
\frac{d P_n(Y)}{d Y} \,\,&=&\,\,-\Gamma(1 \to 2)\,\,n\,\,P_n(Y)\,\,+\,\,\Gamma(1 \to 2)\,(n -1)\,P_{n 
-1}(Y) \,\nonumber \\
\,& & \,\,-\,\,\Gamma(2 \to 1)\frac{n (n -1)}{2}\,\,P_n(Y) \,\,+\,\,\Gamma(2 \to 1)\frac{n (n 
+ 1)}{2}\,\,P_{n+1}(Y)
\eea
\eq{MARCH} has a simple structure: for every process of dipole splitting or merging we see two 
terms. The first one with the negative sign  describes a decrease of probability $P_n$ due to the 
process of splitting or merging of dipoles. The second term with  positive sign is  responsible for the 
increase of the probability  due to the same processes of dipole interactions.

 The boundary conditions are very simple
\beq \label{GABO}
Z(Y;u = 1)\,\,\,=\,\,\sum_{n = 0}^{\infty}\,\,P_n(Y)\,\,=\,\,1
\eeq
which directly follows from \eq{GAZ} and express the conservation of probabilities, saying that 
the total probability is equal to unity at any value of rapidity.
 
Using the Laplace transform
\beq \label{GAMEL}
Z(Y;u)\,\,=\,\,\int^{a + i \infty}_{a - i \infty}\,\,
\frac{d \,\omega}{2\,\pi\,i}\,Z(\omega;u)\,\,\,e^{\omega Y}
\eeq
 we can reduce \eq{GA1} to the  following equation \footnote{In principle, this equation contains an
 inhomogeneous term (initial condition)
 $Z(Y=0,u)$ , however, we will include the correct initial condition latter, after solving this homogeneous
  equation.  }
\beq \label{GASL}
\frac{\omega}{u (1 - u)}\,Z(\omega; u)\,\,=\,\,-\,\kappa\,Z'_u (\omega; u)\,\,+\,\,Z^{"}_{u,u}\, (\omega; u)
\eeq
Introducing new functions
\beq \label{GANF}
s(u)\,\,\equiv\,\,\frac{1}{u (1 - u)}\,e^{- \kappa\,u}\,;\,\,\,\,\,\mbox{and}\,\,\,\,\,
p(u)\,\,\equiv\,\,\,e^{- \kappa\,u}\,;
\eeq
we can reduce \eq{GASL} to the typical Sturm-Liouville equation \cite{KAMKE,POLY}
\beq \label{GASL1}
-\,s(u)\,\omega\,Z (\omega,u)\,\,+\,\frac{d}{d u} \left(p(u)\,Z_u(\omega,u) \right)\,\,=\,0
\eeq
\subsection{General properties of the  Sturm-Liouville problem}
It is very instructive to remind ourselves of the general properties of the Sturm-Liouville problem which has been 
studied by mathematicians for a long time \cite{KAMKE,POLY}.
\begin{enumerate}
\item \quad \eq{GASL1} has the infinite set of the eigenvalues \,$\omega_n = - \lambda_n$.  All $\lambda_n$
are real and can be ordered so that $\lambda_1 \,<\,\lambda_2\,<\,\lambda_3$ with $\lambda_n \,\to\,+ 
\,\infty$. Therefore, there can exist only a finite number of negative $\lambda_n$. But for our particular case, 
(for \eq{GASL1}) there are no negative eigenvalues $\lambda_n$ and $\lambda_1 = 0$ is the least
eigenvalue, to which there corresponds the eigenfunction $Z_0(u) = \mbox{Constant}$;

\item \quad The multiplicity of each eigenvalue is equal to 1;

\item \quad The eigenvalues are determined up to a constant multiplier.

\item \quad  Each eigenfunction $Z_n(u)$ has
exactly $n -1$ zeroes in the interval $(0,1)$;

\item \quad Eigenfunctions $Z_n(u)$ and $Z_m(u)$ are orthogonal with weight $s(u)$, namely,  
\beq \label{GAORT}
\int^1_0\,du\,s(u)\, Z_n(u)\,Z_m(u)\,\,=\,\,0 \,\,\,\,\,\mbox{for}\,\,\,\,\,n \,\neq\,m
\eeq
\item \quad
An arbitrary function $F(u)$, that has a continuous derivative and satisfies the boundary conditions of the
 Sturm-Liouville problem (in other words, a function that we, as physicists, want to find),
can be expanded into absolute and uniformly convergent series in eigenfunctions:
\beq \label{GASER}
F(u)\,\,=\,\,\sum^{\infty}_{n =
1}\,F_n\,Z_n(u)\,;\,\,\,\,\,\mbox{where}\,\,\,\,\,F_n\,\,=\,\,\frac{1}{\|Z_n\|}\,\,\int^1_u\,d 
u'\,F(u')\,Z_n(u')
\eeq
where
\beq \label{GANORM}
\|Z_n\|^2\,\,=\,\,\int^1_0\,d u\,s(u)\,Z^2_n(u)
\eeq

\item \quad The following asymptotic relation holds for large eigenvalues as $n \,\to\,\infty$:
\beq \label{GALN}
\lambda_n\,\,=\,\,\frac{\pi^2\,n^2}{\Delta^2}\,\,\,+\,\,\,O(1)\,;\,\,\,\,\,\,\,\mbox{where}\,\,\,\,\,\,\,
\Delta\,\,=\,\,\int^1_0\,\,
\sqrt{\frac{s(u)}{p(u)}}\,d u\,;
\eeq
\item \quad  In our case (see \eq{GANF}) 
\beq \label{GADELTA}
\Delta\,\,=\,\,\int^1_0\,\frac{d u'}{\sqrt{u' (1 - u')}}\,\,=\,\,\pi,
\eeq
and, therefore, at large $n$ 
\beq \label{GALN1}
\lambda_n\,\,\,=\,\,n^2
\eeq
In  Appendix E we compare this asymptotic spectrum with the spectrum of the exact solution as well as 
with the perturbation procedure 
to calculate it, given in the next section.
\item \quad In out case we can use \eq{GALN1} and \eq{GALN} for $n\,\gg\,1$ to estimate not only the values
of $\lambda_n$ but also to find the corresponding eigenfunctions. These eigenfunctions have the following 
form:
\beq \label{GAEF}
\frac{Z_n(u)}{\|Z_n\|}\,\,=\,\,\left( \frac{4}{\pi^2}\,u\,( 1 -
u)\right)^{\frac{1}{4}}\,e^{\frac{\kappa}{2} \,u}\,\,\sin\left( n\,\Theta(u) 
\right)\,\,+\,\,O\left(\frac{1}{n} \right)
\eeq
where
\beq \label{THETA}
\Theta(u)\,\,=\,\,-\,\int^1_u\,\frac{d u'}{\sqrt{u' \,(1 - u')}}\,\,=\,\,-\,\frac{\pi}{2}\,\,\,-\,\,\,\arcsin( 1 - 2\,u)
\eeq
 \end{enumerate}

Introducing a more general equation \footnote{More useful information on the general Sturm-Liouville 
is given in Appendix A and in Refs.\cite{KAMKE,POLY,RESK}.} than \eq{GASL1}, namely,
\beq \label{GENEQ}
\,s(u)\,\frac{\partial\,Z(Y,u)}{\partial Y}\,\,=\,\frac{\partial}{\partial u} \left( \, p(u) \,\, \frac{\partial \, Z(Y,u)}{\partial u}  \, \right)\,\,+
\Phi(Y,u) 
\eeq
where $\Phi(Y,u)$ is a given function. We can find a 
 general solution to this equation   with the initial condition:
\beq
Z\left(Y=0,u \right)\,\,=\,\,f(u)
\eeq
and with the boundary conditions:
\bea
Z\left(Y,u=0 \right) \,\,\,=\,\,&\,\,\,0; \label{BC1}\\
Z\left(Y,u=1 \right) \,\,\,=\,\,&\,\,\,1; \label{BC2}
\eea

The form of this solution is (see section {\bf 1.8.9} in Ref. \cite{POLY})
\bea \label{GENSOL}
Z\left(Y,u \right)\,\,\,&=& \,\,\,\int^Y_0\,d\,Y'\,\,\int^1_0\,d \xi\,\,\Phi\left(Y',\xi
\right)\,\,G\,\left(Y -Y';u,\xi 
\right)\,\,+\,\,\int^1_0\,d \xi\,\,s(\xi)\,\,f(\xi)\,G\,\left(Y ;u,\xi \right) \nonumber \\
 & + &  \,\,\,p(u=1)\,\int^Y_0\,\,d \,Y'\,\,\,Z\left(Y,u=1 \right)\, \frac{\partial G\,\left(Y - Y' ;u,\xi
\right)}{\partial \xi}|_{\xi =1}\,\,\nonumber \\
 &+&\,\,p(u=0)\,\int^Y_0\,\,d \,Y'\,\,Z\left(Y,u=0 \right) \, \frac{\partial G\,\left(Y - Y' ;u,\xi
\right)}{\partial \xi}|_{\xi =0}
\eea
where the Green function of this  equation is equal to
\beq \label{GRF}
G\,\left(Y ;u,\xi \right)\,\,=\,\,\sum_{n=1}^{\infty}\,\,\frac{Z_n\left( u
\right)\,\,Z_n\left( \xi \right)}{\|Z_n\|^2}\,\,e^{ - \lambda_n\,Y}\,
\eeq
where $Z_n$ are the eigenfunction of \eq{GASL1} with the boundary conditions
\beq \label{BOCONZ}
Z_n(u=0)\,\,\,=\,\,0;\,\,\,\,\,\,\,\,\,\,\,\,\,Z_n(u=1)\,\,\,=\,\,0;
\eeq

One can see that we know a lot about the  properties of the master equation. 

\subsection{The iterative algorithm for  a search of the solution}

The general formula, given in the previous section, leads to a definite algorithm for numerical 
calculations. In the first stage of this algorithm we can use the eigenvalues given by \eq{GALN1}, and 
the eigenfunctions from \eq{GAEF}. The Green function has the form
\beq \label{GF1}
G_{\infty}\,\left(Y ;u,\xi \right)\,\,=\,\,\frac{2}{\pi}\,\,\sqrt{
\sin\Theta(u)\,\,\sin\Theta(\xi)}\,\,e^{ 
\frac{\kappa}{2}\,( u + \xi)}\,
\sum_{n=-\infty}^{n=\infty}\,\,\exp \left( i \,\, n 
\,\,\left\{\,\Theta(u) 
\,\,-\,\,\Theta(\xi)\,\right\}\,\right)\,e^{ - n^2\,Y}
\eeq

The solution can be found from \eq{GENSOL} and it has the form
\beq \label{GF2}
Z_{\infty}\left(Y,u \Lb\Theta \Rb\,\right)\,\,=
\,\int^Y_0\,d\,Y'\,\int^0_{-\pi}\,d
\Theta'\,\, \frac{\frac{1}{2}(\cos \Theta' \,+\,1)}{\sin \Theta'}\,\,e^{ - \kappa u\Lb\Theta \Rb
}\,G_{\infty}\left(Y - Y'
;u\Lb\Theta \Rb
,\xi \Lb\Theta'\Rb\right)
\,\,-
\eeq
$$
  -   \,\,\,\,e^{-\kappa}\,\,
  \int^Y_0\,\,dY'\,\frac{ \partial G_{\infty}\left(Y-Y' ;u\Lb\Theta \Rb,\xi\Lb\Theta'\Rb\right)}{\partial \xi\Lb\Theta'\Rb}|_{\Theta'
   = 0}
$$

In the next stage of the algorithm we are searching for the solution in the form
\beq \label{GF3}
Z\left(Y,u \right)\,\,\,=\,\,\Delta Z\left(Y,u \right)\,\,+\,\,Z_{\infty}\left(Y,u \right)
\eeq
and for $\Delta Z\left(\omega,u \right)$ we obtain the following equation
\beq \label{GF4}
-s(u)\,\omega\,\Delta Z\left(\omega,u \right) \,\, = \,\, L\left(\Delta Z\left(\omega,u
\right)\right)\,\,+\,\,L\left( Z_{\infty}\left(\omega,u \right)\right)
\eeq
where we denote by $L $ the Sturm-Liouville operator, namely $L \,=\,\frac{d}{d u} \left( p(u)
\frac{d}{d u} \right)$.

\eq{GF4} is prepared for the iterative procedure of finding the solution using \eq{GF2} as the first iteration.
To
 calculate the second iteration we substitute $Z_{\infty}\left(Y,u \right)$ into \eq{GF4} which has the form of
 \eq{GENEQ} with $\Phi(Y,u)\,=\,L\left( Z_{\infty}\left(Y,u
\right)\right)$. Using \eq{GENSOL} we can find $ \Delta Z\left(Y,u \right)$.
 Repeating this procedure
several  times we will obtain the numerical solution to the problem. However, the procedure would
be
better converged if we could determine the eigenvalues of the problem at small values of $n$. Below we
suggest two approximations that allows us to approach small values of $n$.

The algorithm described above has a very simple meaning if we re-write our equation as a  Schroedinger- type
equation. It is easy to see that
the search for   $Z_{\infty} \left(Y, u \right)$ of \eq{GF2} can be simplified if we introduce a new
function $T$
\beq \label{T}
Z \left(Y, u \right)\,\,\,=\,\,\, T\left(Y,
\Theta
\right)\,e^{ \frac{\kappa}{2}\,u}
\eeq

For this function we can rewrite \eq{GASL} in the very convenient form for the  function $ T\left(Y,
\Theta\right)$  introducing a new variable $\Theta$ (see \eq{THETA}) instead of $u$.

\beq \label{THETAEQ}
\frac{\omega}{\sin\Theta}\,T\left(\omega, \Theta \right) \,\,=\,\,- \frac{\kappa^2}{4}\,\sin\Theta\,T
\left(\omega, \Theta \right)
\,\,\,+\,\,\,\frac{d}{d \Theta}
\frac{1}{\sin\Theta}\,\frac{d\,\,T \left(\omega, \Theta \right)}{d \Theta}
\eeq
which is a generalized Sturm - Liouville equation of the form

\beq \label{THETAEQ1}
s(\Theta)\,\frac{\partial T \left(Y, \Theta \right)}{ \partial Y}\,\,=\,\frac{\partial}{\partial \Theta}
p(\Theta) \,\frac{\partial\,\,T \left(Y, \Theta \right)}{\partial \Theta}\,\,-\,\,q(\Theta) T \left(Y,
\Theta \right)
\eeq
with
\beq \label{THETAEQ2}
s(\Theta)\,\,=\,\,1/\sin \Theta\,;\,\,\,\,\,\,\,\,\,p(\Theta)\,\,=\,\,1/\sin
\Theta\,;\,\,\,\,\,\,\,\,\,q(\Theta)\,\,=\,\,\frac{\kappa^2}{4}\,\sin\Theta
\eeq
  \eq{THETAEQ1} allows us to find corrections of the order of $1/n^2$ to the spectrum of the
Sturm - Liouville operator. Indeed, $\lambda_n$ is equal to (see section {\bf 1.8.9} in Ref.
\cite{POLY})
\beq \label{THETAEQ3}
\sqrt{\lambda_n}\,\,\,=\,\,n\,\,+\,\,\frac{1}{\pi\,n}\,Q(0,-\pi)\,\,+\,\,O\left( \frac{1}{n^2} \right)
\eeq
where
\beq \label{THETAEQ4}
Q(\Theta, \Theta')\,\,=\,\,\frac{1}{2}\,\,\int^{\Theta}_{\Theta'}\,\,d\,\Theta''
\,\,q(\Theta'')\,\,\,=\,\,\frac{\kappa^2}{8}\,\left\{ \cos\Theta' \,\,-\,\,\cos\Theta \right\}
\eeq
The estimate for the eigenfunctions looks as follows
\beq  \label{THETAEQ5}
T_n(\Theta)\,\,=\,\,\sin( n \Theta) \,\,-\,\,\frac{\kappa^2}{4\,\pi\,n}\,\,\left\{ \,2 \,\Theta + \pi (1
\,-\,\cos\Theta ) \right\} \cos( n \Theta)\,\,+\,\,O\left( \frac{1}{n^2} \right)
\eeq

\FIGURE[h]{
\begin{minipage}{75mm}{
\centerline{\epsfig{file=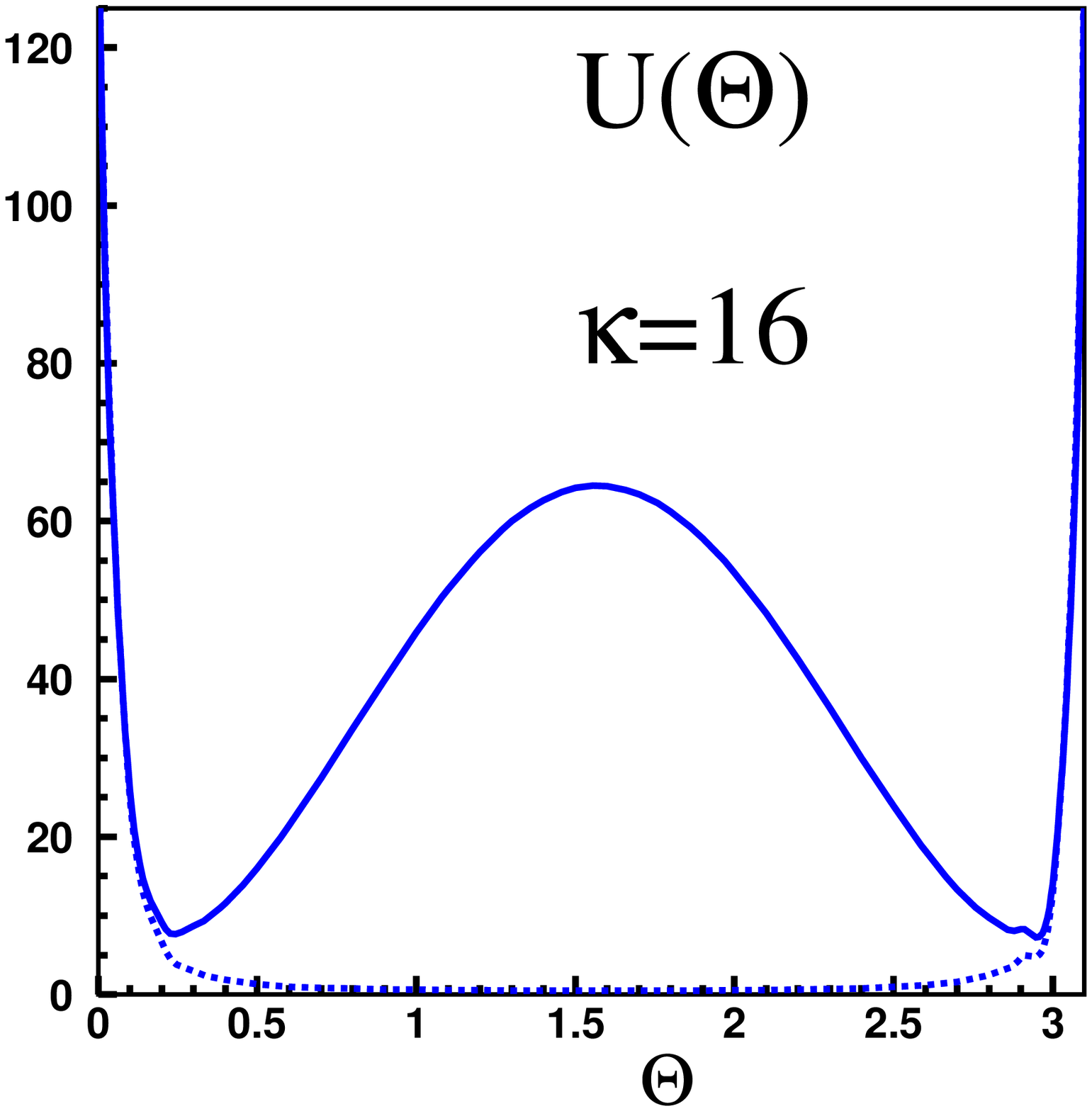,width=80mm}}
}
\end{minipage}
\caption{The potential energy $U \left(\Theta \right)$ of \eq{SCH8} versus $\Theta$ at $\kappa$=16, 
. The dashed line shows the potential $ U \left(\Theta \right)\,\,-\,\,(\kappa^2/4)\,\sin^2 \Theta$ 
which corresponds to the set of the eigenfunctions given by \eq{GAEF}. } 
\label{pot}
}

\eq{THETAEQ} can be re-written as a Schroedinger-type equation:
\beq \label{SCHEQ}
\frac{d^2 \Psi}{(d \Theta)^2}\,\,+\left( E - U\left(\Theta\right) \right)\Psi\,\,\,\,=\,\,\,0\,.
\eeq
Indeed, \eq{THETAEQ} has the equivalent form
\beq \label{SCH6} 
\omega T\,\,=\,\,-\frac{\kappa^2}{4}\,\sin^2 \Theta\,\,T\,\,-\,\,\cot \Theta 
\,\,T_{\Theta}\,\,+\,\,T_{\Theta 
\Theta}
\eeq
Introducing (see Appendix C for general way of reducing the Fokker-Planck equation to the evolution 
described by hermitian Hamiltonian)  
\beq \label{SCH7}
T \,\,=\,\,\phi(\Theta) \,\Psi\left( \Theta \right)\,\,\,\,\,\mbox{with}\,
\,\,\,\,\phi(\Theta)\,\,=\,\,\sqrt{\sin \Theta}
\eeq
we obtain for $\Psi $ the Schroedinger equation with
\beq \label{SCH8}
E\,\,=\,\,-\omega\,\,;\,\,\,\,\,\,\,\,\,\,\,\,\mbox{and}
\eeq
$$
U\left(\Theta\right)\,\,\,\,=\,\,\,\frac{\kappa^2}{4}\,\sin^2 \Theta  \,\,
+ \,\, \frac{1}{4}\,\,(2 + 3 \cot^2 \Theta )\,\,;
$$

In \fig{pot} we plot this potential. One can see that for large $\kappa$ it has
a typical two minima form with the maximum at $\Theta = \pi/2$ while at $\Theta \to 0$ and $\Theta \to
\pi$ \,\,$U(\Theta)$ increases approaching infinity.

 Therefore, for large energy excitation we can replace
this potential by the rectangular-well potential with the wave functions given by \eq{GAEF} with the
spectrum of \eq{GALN}.
Indeed, \fig{pot} shows that $U\left(\Theta  \right) -
\frac{\kappa^2}{4}\,\,\sin^2\left(\Theta\right) \,\,=\,\,
 \frac{1}{4}\,\,(2 + 3 \cot^2 \Theta $)
denoted
as
dashed
line , can be replaced by the potential which is equal to zero for $0 < \Theta < 2\pi$ and which is infinitively large outside of this region.  Such potential allows us
to
 reduce \eq{SCHEQ} to the Schroedinger equation with the simplified potential,
namely,
\beq \label{SCHSIMP}
\frac{d^2 \Psi}{(d \Theta)^2}\,\,+\left( E \,\,-\,\, \frac{\kappa^2}{4}\,\,\sin^2\left(\Theta\right)
\right)\,\Psi\,\,\,\,=\,\,\,0\,\,;\,\,\,\,\,\,\,\,\,\,
\mbox{with}\,\,\,\,\,\,\,\,\,\,\Psi\left(\Theta =0 \right)\,\,=\,\,\Psi\left(\Theta =\pi
\right)\,\,=0\,\,
\eeq
The boundary conditions in \eq{SCHSIMP} reflects the fact that potential is very large outside the $[ 0, 2 \pi]
$ interval in $\Theta$.

 The fact that we have two minima potential leads to the asymptotic behaviour of
the amplitude which is determined by the asymptotic solution of \eq{GASL} for $\omega=0$ (see
Refs.\cite{L4,BOR,AMCP}), namely,
\beq \label{ZASP}
Z_{asymp}(Y = \infty,u)\,\,=\,\,\frac{ 1 - e^{\kappa\,u}}{1 - e^{\kappa}}
\eeq
which satisfy both boundary conditions:  $Z_{asymp}(u =0)\,=\,0$ and $Z_{asymp}(u =1)\,=\,1$ .

In section 4 we will give the exact analytical solution to the master equation and we postpone until  this
section our  discussion of this Schroedinger-type of the equation. However, this equation (see
\eq{SCHEQ}) allows us to use the usual perturbative approach for the numerical solutions.
For example we can represent the potential $U(\Theta)$ of \eq{SCH8} in the form
\beq \label{SCH9}
U\left(\Theta \right)\,\,=\,\,U_0\left(\Theta \right) \,\,+\,\,V\left(\Theta \right)\,\,\,\,\,
\eeq
and solve \eq{SCHEQ} with $U_0$ exactly. Considering $V\left(\Theta \right)$  as being small we can
develop the usual perturbation approach in quantum mechanics for a numerical solution of the equation.
The value of $\kappa$ is large in our approach and we need to take such a perturbation procedure with
a great  precaution. In appendix E we give the spectrum of $\lambda_n$ calculated in the first order of 
preturbation approach and  it terms out the agreement with the exact spectrum is very impressive.

 However for large values of $\kappa$ the more reasonable way
 to find a good first approximation for searching $Z_{\infty}$ is to use the
semi-classical
approach which we consider below.

\section{Analytical solution  to the  Sturm-Liouville problem }
\label{sec:SC}
\subsection{Semi-classical approach}
In our master equation (see \eq{GASL}) we have a natural large parameter: $\kappa$ (see \eq{KAPPA}).
We wish to use this parameter to develop a semi-classical approach to the master equation. We assume that
\beq \label{SC1}
Z(\omega;
u)\,\,\,=\,\,\,e^{\phi(\omega;u)}\,\,\,\,\mbox{where}\,\,\,\,\phi^{"}_{u,u}\,\,\,\ll\,\,(\phi^{'}_{u})^2
\eeq
Substituting \eq{SC1} into \eq{GASL} one can reduce this equation to the form
\beq \label{SC2}
\frac{\omega}{u (1 - u)}\,\,=\,\,-\kappa\,\phi^{'}_{u}\,\,\,+\,\,\,(\phi^{'}_{u})^2
\eeq
Solving \eq{SC2} we obtain
\beq \label{SC3}
\frac{d \phi^{\pm}_u}{d u}\,\,\,=\,\,\,\frac{\kappa}{2}\,
\left(\,\,1\,\,\pm\,\,\,\sqrt{1\,\,+\,\,\frac{4\,\omega}{\kappa^2\,u\,(1
- u)}}\,\,\,\,\,\,\, \right)
\eeq
and 
\beq \label{SC4}
\phi^{\pm}\left(\omega; u\right) \,\,\,\,=\,\,-\,\int^1_u\,\,\phi^{\pm'}_{u'} (\omega; u')\,\,d\,u'
\eeq
The general solution to \eq{SC1} has the form
\beq \label{SC5}
Z(Y;u)\,\,\,=\,\,\,\int ^{a + i \infty}_{a - i \infty}\,\,\frac{d \omega}{2\,\pi\,i}\,e^{ 
\omega\, Y}\,\,\left\{
\Phi^{(-)}(\omega)\,e^{ \phi^{-}(\omega,u)}\,\,\,+\,\,\Phi^{(+)}(\omega)\,e^{ \phi^{+}(\omega,u)}
\right\}
\eeq
where functions $\Phi^{(-)}(\omega)$ and $\Phi^{(+)}(\omega)$ should be found from initial and boundary
conditions (see \eq{GAIN} and \eq{GABO}).  We recall that we use  $Y$ for ${\cal Y} = \Gamma(2 \to 1)\,Y$.

The contour of integration (contour $C$ in \fig{contr})   over $\omega$ is situated to the right of
the singularities as it is shown in 
\fig{contr}. As we have discussed the general structure of the singularities looks as follows. We have no
 singularities for positive $\omega = \omega^+_n > 0$. It looks strange since one can see from \eq{SC3} 
that we have a restricted number of singularities for positive $\omega\, = \, \omega^+_n\,>\,0$. 
However, 
since the 
contour  of integration $C$ in \fig{contr} is placed  between $\omega_0\,=\,0$ and $\omega =
\omega^+_1$, we
can close it to the left semi-plane (contour $C_1$ in  \fig{contr}) and, therefore, the singularities
$\omega^+_n$ does not contribute to the integral of \eq{SC5}.

\FIGURE[h]{
\begin{minipage}{85mm}{
\centerline{\epsfig{file=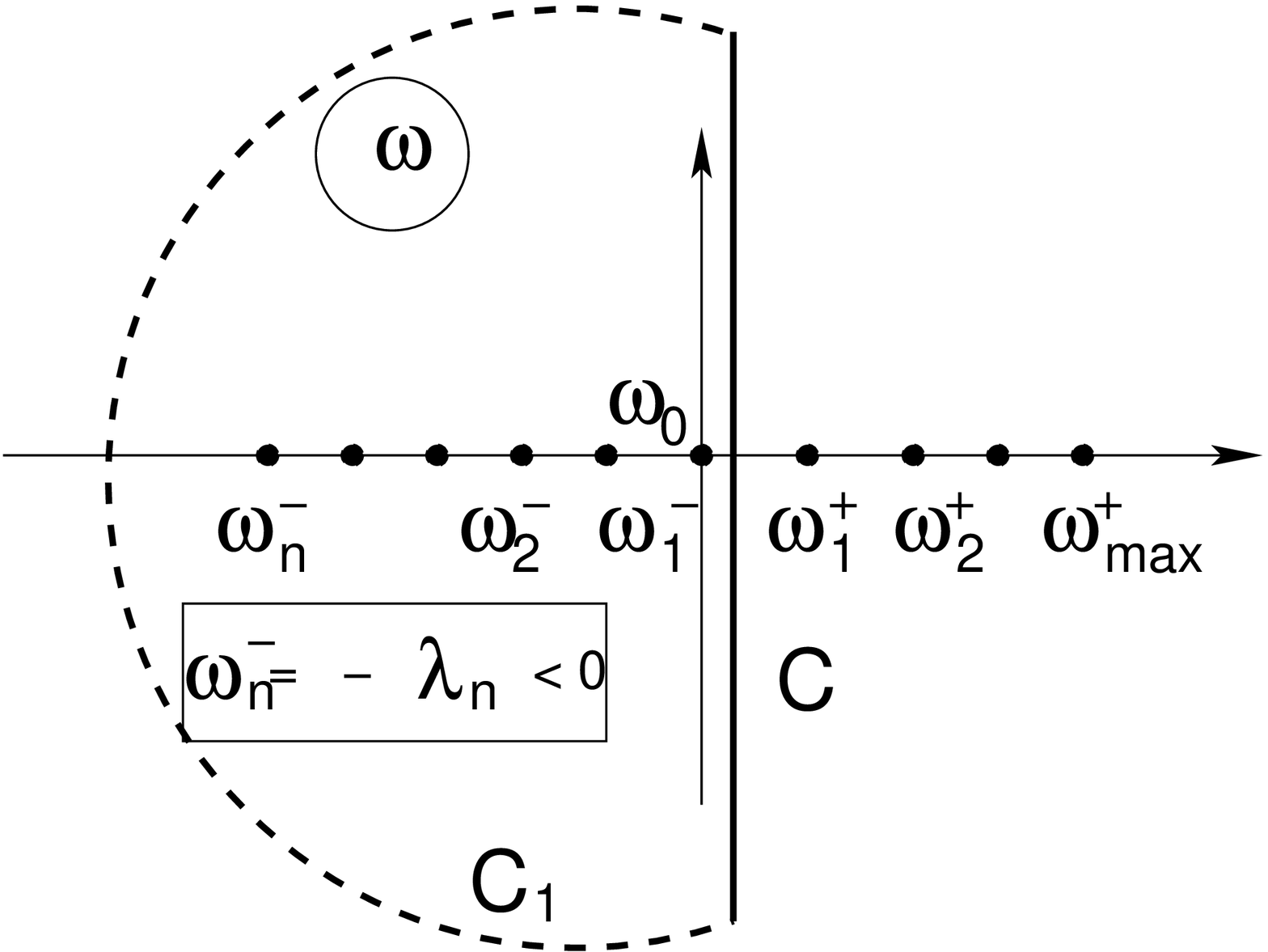,width=80mm}}
}
\end{minipage}
\caption{The contours of integration over $\omega$ in \eq{SC5}. All singularities are located along real
 negative real axis ($\omega_n = - \lambda_n \,< \,0 $)}
\label{contr}
}

The analytical function for \eq{SC4} is written in the Appendix B  while here we will discuss this
function in
two different limits: $\frac{\omega}{\kappa^2 u (1 - u)}\,\,\ll\,\,1$ and   $\frac{\omega}{\kappa^2 u (1 -
u)}\,\,\gg\,\,1$.

For $\frac{\omega}{\kappa^2 u (1 - u)}\,\,\ll\,\,1$ we have for function $\phi^{\pm}\left(\omega; u\right)$
the following expression
\bea
\phi^{+}\left(\omega; u\right) \,\,\,\,&=& \,\,\,\kappa( u - 1) \,\,+\,\,\frac{\omega}{\kappa}\,\ln
\left(\frac{u}{ 1 - u} \right);\label{SC6+} \\
\phi^{-}\left(\omega; u\right) \,\,\,\,&=& \,\,\, \,\,-\,\frac{\omega}{\kappa}\,\ln
\left(\frac{u}{ 1 - u} \right); \label{SC6-}
\eea

It is easy to see that $\phi^+(\omega;u)$ of \eq{SC6+}  corresponds to the solution which survives at $\kappa
\,\to\,\infty$
which describes the case of the mean field approach. Analyzing the mean field solution (see Ref. \cite{L1} ,
for example) one can see that all positive singularities $\omega^+$ in \fig{contr}) are the singularities of
$\Phi^{(-)}$ while $\Phi^{(+)}$ has the singularities at the negative values of $\omega$. Therefore,
the only singularities that we need to take into account are those that are described by \eq{SC6+}.

For $\omega \approx \kappa^2 $ we need to expand the exact formulae for $\phi^{\pm}(\omega,u)$ in
 the kinematic
region. \,\,\,\,        For $\frac{\omega}{\kappa^2 u (1 - u)}
\,\,\gg\,\,1$.
In this region  $\phi^{\pm}(\omega,u)$ are equal to
\beq \label{SC7}
\phi^{\pm}(\omega,u)\,\,\,=
\eeq
$$
\,\,\,\frac{\kappa}{2}( u  - 1 )\,\,\pm\,\,i \,\sqrt{|\omega|}
\,\,\Theta(u)\,\,=\,\,\frac{\kappa}{2}( u - 1)\,\,\pm\,\,i \,\mu_n \,\,\Theta(u)
$$
where $\Theta $ is given by \eq{THETA} and
$\omega_n = - \lambda_n = - \mu^2_n$.

Therefore, in both branches $\phi^{\pm}$ we have singularities at negative $\omega$ and from estimates of
\eq{GALN1} we know that $\lambda_n = n$.
Using simple relations between variables $u$ and $\Theta$
\beq \label{SCKR}
u\,\,=\,\,\frac{1}{2}\,\left( \cos \Theta\,\,+\,\,1 \right)  \,;
 \eeq
we can rewrite
 $\exp\left( \phi^{\pm} (\omega, u) \right)$ in the form
\beq \label{SC8}
\exp \left( \phi^{\pm} (\omega_n = - \lambda_n = - \mu^2_n, u)
\right)\,\,\,=\,\,\,\exp\left(\frac{\kappa}{8}\,\left[
-2\,+ \zeta
\,+\,\frac{1}{\zeta} \right] \right)\,\zeta^{\pm\,\mu_n}
\eeq
where we introduce new variable
\beq \label{ZETA}
\zeta\,\,\,=\,\,e^{i\,\Theta}
\eeq

It is obvious from \eq{SC7} that $\zeta^n$ is the eigenfunction of our master equation at large values
of $\omega$ \,\,\,\,\,\,($\omega \geq \kappa$). However, the eigenfunctions that satisfy the boundary
conditions
of \eq{BOCONZ}  are
\beq \label{ZNSIN}
Z_n\left( \Theta \right)\,\,=\,\,\sin\left( n\,\Theta \right)
\eeq

These functions form the complete set of functions in the interval $0 \,\leq\, u \,\leq\, 1$.
In other words, the approximation of  \eq{GALN1} turns out to be exact in our semi-classical approach.
Having this fact in mind,  we can  find function $\Phi^{(\pm)}(\omega) $ in \eq{SC5} from
 the initial condition of \eq{GAIN}.

Functions $\zeta^n$ are eigenfunctions of \eq{THETAEQ} in which we neglect the first term in the r.h.s.
and the second one is modified assuming that
  $$\frac{d}{d \Theta}
\frac{1}{\sin\Theta}\,\frac{d\,\,T \left(\omega, \Theta \right)}{d \Theta}\,\,\, \mbox{~ is replaced
by ~}\,\,\,
\frac{1}{\sin\Theta} \, \frac{d^2\,\,T \left(\omega, \Theta \right)}{d \,\Theta^2} .$$

In this approximation  \eq{THETAEQ} reduces to
\beq \label{THETAEQ6}
\omega \,T\left(\omega, \Theta \right) \,\,=\,\,
\,\,\,\,\,\frac{d^2\,\,T \left(\omega, \Theta \right)}{d \Theta^2}
\eeq
 It is easy to see that \eq{ZNSIN}
  gives the eigenfunctions of this equation.

  Using the solution of \eq{THETAEQ6} we can calculate the Green function for our master equation (see \eq{GASL})
that this
Green
 function  has a very simple form, namely
\beq \label{THETAEQ7}
G^{SC}_{\infty}\,\left( Y; \Theta, \Theta'
\right)\,\,=\,\,\frac{1}{\pi}\,e^{\frac{\kappa}{2} +
\frac{\kappa}{4}\,[\cos \Theta \,+\,\cos \Theta' ]}\,\,\,\sum^{\infty}_{n =1}\,\,
\sin\left(\,n\,\Theta \right)\,\,\sin\left(\,n\,\Theta' \right)\,\,e^{ - n^2\,Y}
\eeq
and the initial conditions have the form
\beq \label{INT}
T\left(Y=0,\Theta \right)\,\,\,=\,\,\frac{1}{2} ( 1 + \cos\Theta)\,\exp \left( - \frac{\kappa}{4}\,(1 +
\cos\Theta) \right)
\eeq
Using this Green function we obtain the  semi-classical solution in the form
\bea \label{THETAEQ8}
Z^{SC}_{\infty}\,\,&=&\,\,\\
  & & \frac{1}{\pi}\,\,\sum^{\infty}_{n = -\infty}\,
\,\left\{\,e^{ - n^2 Y}\,\,e^{\frac{\kappa}{2}\,u}\,\,\sin \left(n\,\Theta \right)\,\,\int^0_{-\pi}
\,\,d\,\Theta'\,\sin \left(n\,\Theta' \right)\
 \,\frac{1}{2}\,( 1 + \cos \Theta')\,\exp \left( - \,\frac{ \kappa}{4}\,( 1 + \cos\Theta')
\right)\,\right\}\,\nonumber
\eea

The integral over $\Theta'$ could be taken using the generating function for the modified Bessel
function  of the first kind (see formula {\bf 9.6.33} in Ref.\cite{AS})
\beq \label{SC9}
\exp\left( \frac{\kappa}{4}\,\cos \Theta \right)\,\,=\,\,\sum^{\infty}_{n = - \infty}\,\,\,I_n
\left(\frac{\kappa}{4}\right)\,\,e^{i\,n\,\Theta}
\eeq

and the expression for simple integral
\beq \label{SC91}
\int^0_{- \pi}\,d\,\Theta'\,\,\cos \left( n \Theta' \right)\,\,\sin\left( k \Theta' \right)
\,\,=\,\,-\,\frac{k\,( 1 - (-1)^{n + k})}{k^2 - n^2}
\eeq

From \eq{SC9} and \eq{SC91} one can derive that
\beq \label{SC10}
C( n,\kappa)\,=\,\int^0_{-\pi}
\,\,d\,\Theta'\,\,\sin( n \Theta')
 \,\frac{1}{2}\,( 1 + \cos \Theta')\,\exp \left( -\frac{ \kappa}{4}\,( 1 + \cos\Theta') 
\right)\,
\,\,
=\,\,\frac{1}{2} \exp\left(-\kappa/4 \right)\,\times
\eeq
$$
\left\{\sum^{\infty}_{m=1}\,\left(\,I_{m + 1}( -\kappa/4)
\,+\,I_{m - 1}(\,
-\kappa/4)\,+\,
2\,I_{m}( -\kappa/4)\,\right)\,\,\frac{n\,( 1 - (-1)^{n + m})}{n^2 - m^2}\,\,+ \,\,\left(\,I_{1}( -\kappa/4)
\,\,+\,\,
2\,I_{0}( -\kappa/4)\,\right)\frac{(-1)^n - 1}{n}\, \right\}
$$
Substituting \eq{SC10} into \eq{THETAEQ8} we obtain
\beq \label{SC11}
Z^{SC}_{\infty}\,\,=\,\,\frac{1}{\pi}\,\,e^{ \frac{k}{2}(u -1)}\,\left( C(0,k)\,\,+\,\,2\sum^{\infty}_{n 
= 1}\,
\,\left\{\,e^{ - n^2 Y}\,\,e^{\frac{\kappa}{2}\,u}\,\,\sin(n\,\Theta)\,C(n,\kappa) \right\}\right)
\eeq

However, this expression for $Z^{SC}_{\infty}$ has two disadvantages: (i) we approximate $\omega_n = - 
n^2$ even at small values of $n$ where we know that our spectrum is different (see \eq{SC6+} and 
\eq{SC6-}); and (ii) $Z^{SC}_{\infty}(Y,u=1)\,\neq \,1$ (actually $Z^{SC}_{\infty}(Y,u=1) =0$).

The set of eigenfunction at small values of $n$ is clear from \eq{SC6+} and \eq{SC6-}, namely,
\beq \label{ZN}
Z_n(u)\,\,=\,\,\left\{ \begin{array}{l l  l} \,\,\mbox{for}\,\,\,\,\, u \,\geq \frac{1}{2}\,
&\,\,\,\,\,\,\,&\,\,\,e^{ \kappa\,
(u
-1)}\,\tau^n \,\,=\,\,\exp \left(
\kappa\,\frac{\tau}{1 - \tau}
\right)\,\tau^n\,\,\,\mbox{where}\,\,\tau\,\,=\,\,\frac{u -1}{u}  \\ \\ \\
 \,\,\mbox{for}\,\,\,\,\, u \,\leq\, \frac{1}{2}\,&  & \,\,\,\tau^n
\,\,\,\mbox{where}\,\,\tau\,\,=\,\,\frac{u}{u-1}\end{array}
\right.
\eeq

The fact that we have two different sets of the eigenfunctions for $u \geq 1/2$ and  $u \leq 1/2$,
directly  comes
 from the contour integral of \eq{SC5} and the expressions of \eq{SC6+} and \eq{SC6-} for
functions $\phi^+$ and $\phi^-$. Indeed, for $u \geq 1/2$ $\ln\left(u/(1 - u)\right)$ is positive and we
can close the  contour in \eq{SC5} to the left semi-plane for  $\phi^+$.  For  $u \leq 1/2$ this log is
negative and we have to use $\phi^-$ for negative values of $\omega$. Recall, that we know from the
general properties of the Sturm-Liouville equation that only negative $\omega$ contribute.

The value of $ \omega_n = - \lambda_n$ for the eigenfunctions of \eq{ZN} is equal to
\beq \label{SOM}
\lambda_n\,\,\,=\,\,\,\kappa\,\,n
\eeq
Coefficients $C(n,\kappa)$ we can find using the following series
\beq \label{SC12}
u \,\exp\left(- \kappa ( u - 1) \right)\,\,=\,\,\frac{1}{ 1 - \tau}\,\exp
\left(- \kappa \frac{\tau}{\tau - 1} \right)\,\,=\,\,\sum^{\infty}_{n=0}\,\,L_n(-\kappa)\,\tau^n
\eeq
where $L_n(-\kappa)$ is the  Laguerre polynomial (see  formula {\bf 8.975} in     Ref.\cite{RY}).

From \eq{SC12} one can see that the initial condition $Z(Y=0,u) = u$ can be re-written in the form of
\beq \label{SC121}
u\,\,=\,\,\sum^{\infty}_{n =0}\,\,C(n,\kappa)\,Z_n\Lb u  \Rb\,\,\Theta \Lb u\,\geq \frac{1}{2} \Rb
\eeq
with   $C(n,\kappa)\,\,=\,\,L_n(-\,\kappa)$.

We suggest to build the Green function  considering for all $\lambda_n \leq \kappa^2$ the set of the
eigenfunction given by \eq{ZN} while for $\lambda_n >  \kappa^2$ we use the set of \eq{ZNSIN}.

It means that our Green function has the form
\bea \label{SCGRF1}
G^{SC}\left(Y; u,\xi \right)\,\,&=&\,\,\exp\left(\kappa\,(u + \xi - 2)\right)\,\,\sum^{n=[\kappa] +
1}_{n=0}\,\tau(u)\,\tau(\xi)\,e^{- n\,\kappa Y}\,\,\nonumber\\
 &+&\,\,\exp\left(\frac{\kappa}{2}\,(u + \xi - 2)\right)
\,\,\sum^{\infty}_{n = [\kappa]+2}\,\sin\left(n\,\Theta(u)\right)\,\sin\left(n\,\Theta(\xi)\right)\,\,e^{- n^2\, Y}
\eea
which leads to the generating function in the form

\bea \label{SCGRF2}
Z^{SC}\left(Y; u \right)\,\,&=&\,\,\,\Theta \left( u \,\,\geq\,\,\frac{1}{2}
\right)\,\,\,\exp\left(\kappa\,(u  - 1)\right)\,\sum^{n=[\kappa] +
1}_{n=0}\,\tau(u)^n\,L_n(-\kappa)\,\,e^{- n\,\kappa Y}\,\,  \\
 &+&\,\,\Theta \left( u \,\leq\,\frac{1}{2}
\right)   \,\,\sum^{n=[\kappa] +1}_{n=0}\,\tau(u)^n \,\,e^{- n\,\kappa Y}
 \,\,+\,\,\exp\left(\frac{\kappa}{2}\,(u  - 1)\right)
\,\,\sum^{\infty}_{n = [\kappa]+2}\,\sin\left(n\,\Theta(u)\right)\,C(n,\kappa)\,\,e^{ - n^2 \,Y} \nonumber
\eea

This solution has two problems which cannot be solved analytically: a dependence of the separation
 parameter between large and small $n$ which was taken in \eq{SCGRF2} to be equal to $[\kappa]+1$ and
rather complicated expression for $C(n,\kappa)$ (see \eq{SC10}) which  does not allow us to make all
calculation analytically. We investigated both problems computing \eq{SCGRF2} numerically for
different
separation parameters. It turns out that the answer is very insensitive to the exact value of the
separation parameter if only we take it around $\kappa$.  Actually, the value of this parameter we can
find comparing exact semi-classical functions $\phi^{\pm}$ with our approximate ones ( see Appendix
B,
where we discuss the exact functions).

 The behaviour of the scattering amplitude
is illustrated in \fig{asympy} as a function of rapidity $Y$. In  \fig{asympu} one can see the $u$
dependence of our solution at large values of rapidity $Y$. This behaviour confirms the expectation that
 has been discussed in Ref. \cite{L4}, that at high energy our master equation predicts the gray disc
behaviour for the scattering amplitude.

We check the accuracy of the semi-classical solution by  calculating the following function:
\beq \label{ERROR}
\Delta(Y,u)\,\,=\,\,\frac{\left( \frac{\partial}{\partial Y}\,\,+\,\,\kappa\,u\,(1 - u)\,\,
\frac{\partial}{\partial u}\,\,-\,\,u\,(1 - u)\,\,\frac{\partial^2}{(\partial u)^2}\,\right)\,Z^{SC}\left(
Y,u \right)}{\mbox{minimal separate term of the numerator}}
\eeq

As we have expected the semi-classical approach is rather bad at small values of $Y \leq\,0.05$ leading to
$\Delta(Y,u)\,\,\approx\,\,40\%$ for $u\,\approx 0.5$. However, for large values of $Y$ ($Y \,>\,0.1$
the accuracy is better than 10 \% for all values of $u$. It should be stressed that for $u >0.8$ in the
entire kinematic region of $Y$ the accuracy is not worse than $10\%$. All estimates were done for $\kappa =
16$.  This fact is very encouraging since the scattering amplitude $N$ can be calculated in the following
way \cite{BK,L2}
\beq \label{AMP}
N \left(Y, \gamma(Y_0)\right)\,\,=\,\,1\,\,\,-\,\,\,Z \left(Y, u = 1 -  \gamma(Y_0)\right)
\eeq
where $\gamma (Y_0)$ is the scattering amplitude at low energy ($Y$=0). As it is discussed in Refs. \cite{BK,L2} in \eq{AMP} we assume that $n$ dipoles interact with the target independently. It means that the scattering amplitude
of $n$ dipoles with the target $\gamma_n(Y_0) =  \gamma^n(Y_0)$. \footnote{This contribution for low energy interaction of $n$ dipoles with target is proven in ~Ref.\cite{BRAN}. He has also been used in the mean field approximation ~\cite{BK}. Nevertheless, it looks as an additional assumption for dipole-dipole scattering. We thank Al.Mueller and E.Iancu for very fruitful discussions on this subject.}
 In our QCD motivated model $ \gamma
\,\,\approx\,\,1/\kappa \,\,\ll\,\,1$. Therefore, we are  interested only in the values of $u$ that are
close to unity ($u \to 1$), where the semi-classical approach works quite well.

\DOUBLEFIGURE[ht]{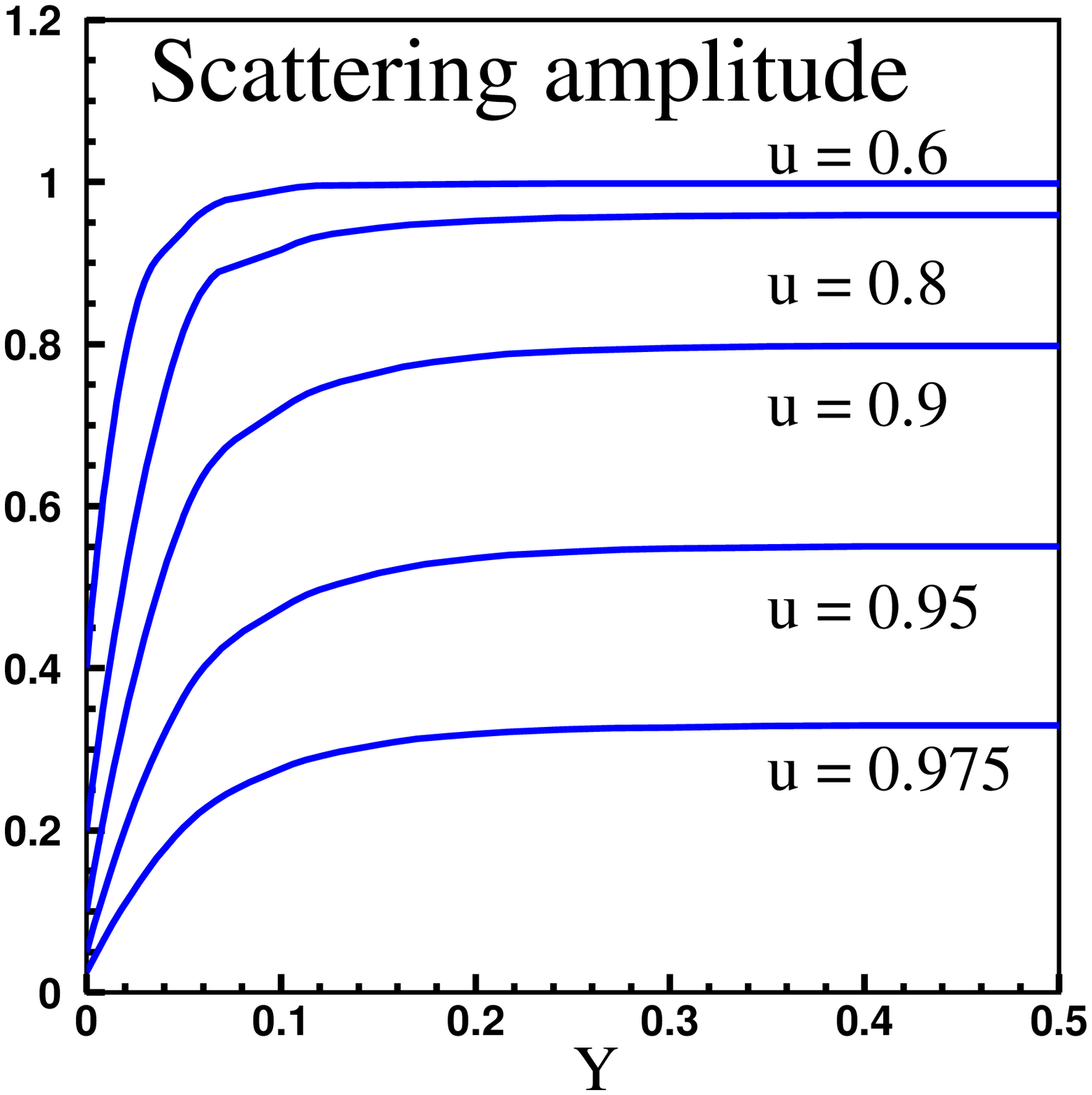,width=75mm}{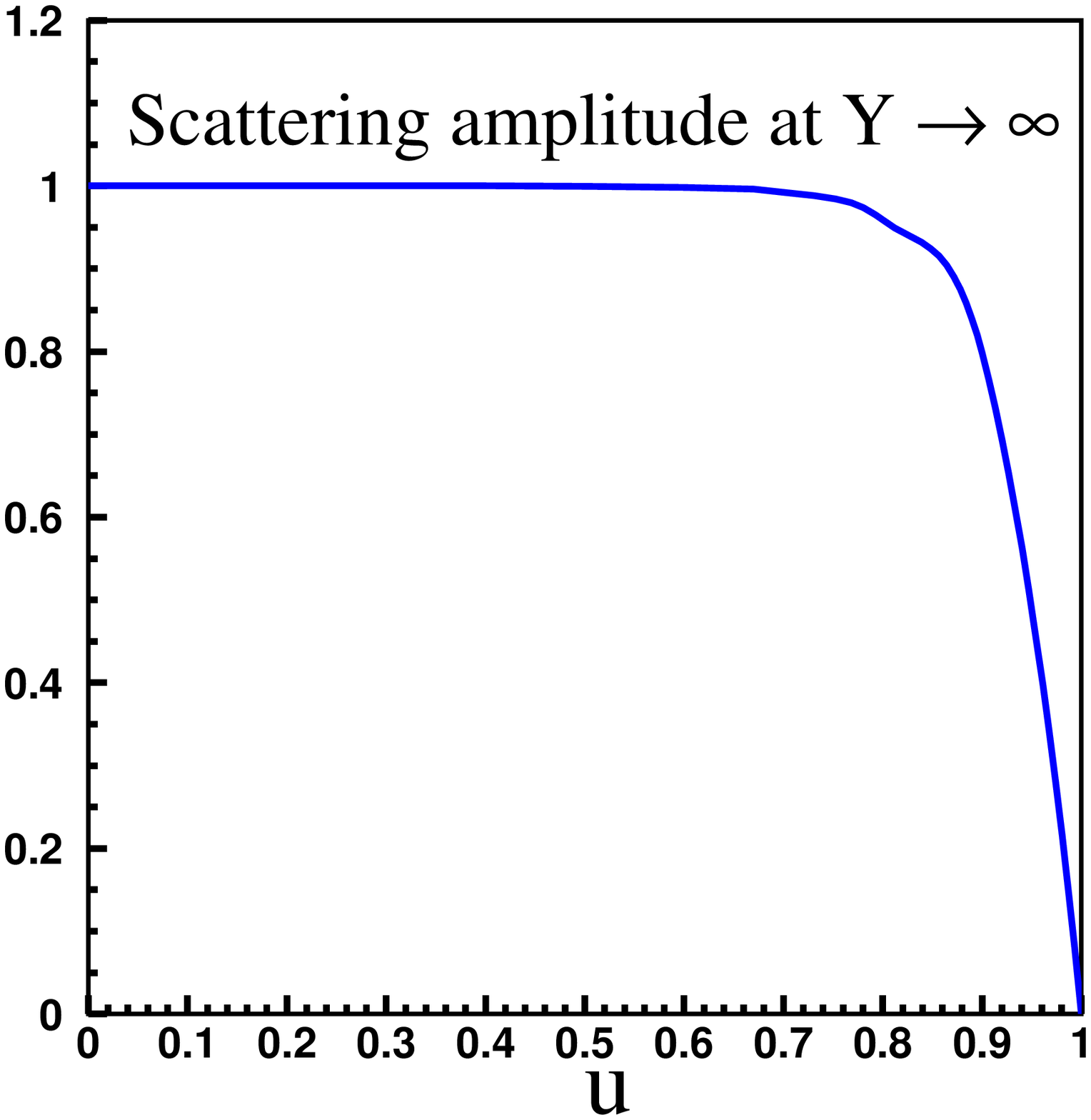,width=75mm}{The
dependence of the scattering
amplitude versus rapidity at fixed $\kappa=16$. Recall that
here  we denote $Y \equiv {\cal Y}$ (see \protect\eq{CALY}.
 \label{asympy}}{The asymptotic dependence of the scattering amplitude (see
\protect\eq{SCGRF2}
)
versus $u$ at fixed
$\kappa=16$.
\label{asympu}}

However, in the next section we develop a better analytical approach for such $u$.

\begin{boldmath}
\subsection{Exact solution at $u \to 1$}
\end{boldmath}

To check that the eigenvalues in the region of small $\omega$ is determined by \eq{SOM} we develop a
different analytical approach. In the vicinity of $u =1$ we can replace \eq{GA1} by the following
equation:
\beq \label{AU1}
\frac{\partial Z}{\partial{\cal Y}}\,\,=\,\,\,(1 - u)\,\left(-\,\,\kappa\,\,
\frac{\partial
Z}{\partial
u}\,\,+\,\,u\,\,
\frac{\partial^2 Z}{(\partial u)^2} \right)
\eeq
For $Z\left(\omega,  u \right)$ \eq{AU1}   reduces to the form
\beq \label{AU2}
\omega\,Z\left(\omega,  u \right)\,\,=\,\,-\kappa\,(1 - u)\,\,Z_u\left(\omega,  u \right)+\,\,u (1 -
u)\,\,Z_{u u}\left(\omega,  u \right)
\eeq

The solution to \eq{AU2} is the  Gauss's hypergeometric function
\beq \label{AU21}
Z\left(\omega,  u \right)\,\,\,=\,\,\,
u^{1 + \kappa}\,\,{}_2F_1\left(\alpha,\beta,\gamma,u
\right)
\eeq
(see formula {\bf 9.151} in Ref. \cite{RY}) with the following parameters
\bea \label{AU3}
\alpha\,\,&=&\,\,\frac{1}{2}\,(1 \,+\,\kappa)\, \,\,+\,\,\,\frac{1}{2}\,\sqrt{(1
\,+\,\kappa)^2\,-\,4\,\omega}\,\,; \nonumber \\
\beta\,\,&=&\,\,\frac{1}{2}\,(1 \,+\,\kappa)\,
\,\,-\,\,\,\frac{1}{2}\,\sqrt{(1
\,+\,\kappa)^2\,-\,4\,\omega}\,\,; \nonumber \\
\gamma\,\,&=&\,\,\kappa + 2\,\,;
\eea
The generating function that satisfies the boundary condition $Z\left(Y,u=1\right)\,=\,1$ is equal to
\beq \label{AU4}
Z\left(\omega,  u \right)\,\,\,=\,\,\,u^{1 +
\kappa}\,\frac{1}{\omega}\,\frac{{}_2F_1\left(\alpha,\beta,\gamma,u
\right)}{{}_2F_1\left(\alpha,\beta,\gamma,u=1
\right)}
\eeq
Since (see formula {\bf 9.122} in Ref. \cite{RY})
\beq \label{AU5}
 {}_2F_1\left(\alpha,\beta,\gamma,u=1
\right)\,\,=\,\,\frac{\Gamma\left(\gamma \right)\,\Gamma (1)}{\Gamma( \gamma - \alpha)\,\,\Gamma( \gamma
- \beta)}
\eeq
The eigenvalues stem from the poles of $\Gamma( \gamma - \alpha)$ or $ \Gamma( \gamma
- \beta)$, namely, from the  equation
\beq \label{AU6}
\gamma - \alpha\,\,=\,\,-n\,\,;\,\,\,\,\,\,\,\,\mbox{which
gives}\,\,\,\,\,\,\,\,\omega_n\,\,=\,\,-\lambda_n\,\,=\,\,-n^2
\,\,-\,\,n\,( 1 + \kappa )\,\,\,\,\,\mbox{where}\,\,\,n\,\,=\,\,1,2,3, \dots
\eeq
This equation leads to a smooth transition between large and small values of $n$ supporting our
semi-classical formula.  Substituting \eq{AU4} in \eq{SC5} one can find the eigenfunction of our
equation. They have the following form
\beq \label{AU7}
Z_n(u)\,\,\,=\,\,\frac{1}{\|Z_n\|^2}\,\,u^{1 + \kappa }\,\,\,P^{-1,1 + \kappa}_n
\left(2\,u\,-\,1 \right)
\eeq
where $P^{-1,1 + \kappa}_n $ are the Jacobi polynomials  (see  sections {\bf 8.960 -
8.967} in Ref. 
\cite{RY}). 

The normalization of function $Z_n$ in \eq{AU7} is obvious since
\beq \label{AU8}
\|Z_n\|^2\,\,=\,\,
\eeq
$$
   \int^1_0\,d\,u\,\,\,\frac{u^{1 + \kappa}}{ 1 -u}\,\,\left( P^{-1,1 + \kappa}_n 
\left(2\,u\,-\,1 \right) \right)^2
\,\,=\,\,\frac{\Gamma \left(n \right)\,\,\Gamma \left(n + 2 + \kappa\right)}{n!\,(1 +\kappa + 
2\,n)\,\Gamma \left(1 + \kappa + n \right)}\,\,=\,\,\frac{\left(n + 1 + 
\kappa\right)}{n\,(1 +\kappa +2\,n)}
$$

It is easy to check that eigenfunctions of \eq{AU7} satisfy the following boundary conditions;
\beq \label{AU9}
Z_n(u=0)\,\,\,=\,\,\,0\,\,;\,\,\,\,\,\,\,\,\,\,\,\,\,Z_n(u=1)\,\,\,=\,\,\,0
\eeq

We found that the easiest way to solve the master equation (see \eq{AU1}) is to introduce $Z(Y,u) 
\,=\,u\,+\,\tilde{Z}(Y,u)$. The equation for $\tilde{Z}$ can be  reduced to the Sturm-Liouville form, namely,
\beq \label{AU10}
s(u)\frac{\partial 
\tilde{Z}}{\partial\,Y}\,\,\,=\,\,
\frac{\partial}{\partial\,u}\,p(u)\,\,\frac{\partial\,\tilde{Z}}{\partial\,u}\,\,+\,\,\Phi(u)
\eeq
where
\beq \label{AU11}
s(u)\,\,=\,\,\frac{1}{u ( 1 - 
u)}\,u^{-\kappa}\,\,;\,\,\,\,\,\,\,p(u)\,\,=\,\,u^{-\kappa}\,\,;\,\,\,\,\,\,\,
\Phi(u)\,\,=\,\,-\kappa\,u^{-\kappa}
\eeq

The solution to \eq{AU10} has the form
\beq \label{AU12}
\tilde{Z}(Y,u)\,\,\,=\,\,\,\,\,\int^y_0\,\,d\,Y'\,\int^1_0\,\,d\,\xi\,\,G\left(Y -
Y'; u, \xi\,\right)\,\,\Phi (\xi)
\eeq
where
\beq \label{AU13}
G\left(Y ; u, \xi\,\right)\,\,\,=\,\,\sum^{\infty}_{n=1}\,\,\frac{Z_n (u)\,\,Z_n(\xi)}{\|Z_n\|^2} \,\,\exp\left( -\,\lambda_n\,Y \right)
\eeq
where $\lambda_n$ are given by \eq{AU6} and $Z_n$ is the set of the eigenfunction shown in \eq{AU7}.

In \fig{u1} we  plotted the energy behaviour of the scattering amplitude for different values of $u$ that
are close to unity. One can see that this solution gives the amplitude which is quite different from the
semi-classical solution. It sounds controversial, but actually it is not, since the solution of
\eq{AU12} contains the integral over $\xi$ from 0 to  1 while our functions $Z_n$ of \eq{AU7} we
know
only at $u \to 1$.

It turns out that we can find a very simple solution  at small values of $u$ ($u \to 0$).
Introducing a new variable
$\gamma = 1 - u$ we can rewrite our master equation in the form at $\gamma \to 1$
\beq \label{AU14}
\frac{\partial Z}{\partial{\cal Y}}\,\,=\,\,\,(1 - \gamma)\,\left(\,\,\kappa\,\,
\frac{\partial
Z}{\partial
\gamma}\,\,+\,\,\gamma\,\,
\frac{\partial^2 Z}{(\partial \gamma)^2} \right)
\eeq
or in $\omega$-representation it looks as
\beq \label{AU15}
\omega\,Z(\omega,\gamma)\,\,=\,\,(1 - \gamma)\,\left( \kappa\,\,Z_{\gamma}(\omega,\gamma)
\,\,+\,\,\gamma\,Z_{\gamma \gamma}(\omega,\gamma) \right)
 \eeq

The solution to this equation has the form
\beq \label{AU16}
Z(\omega,\gamma)\,\,\,=\,\,\,{}_2F_1\left(\alpha, \beta, \kappa,\gamma
\right)\,\,\,=\,\,\,{}_2F_1\left(\alpha, \beta, \kappa, 1 - u \right)
\eeq
where
\bea \label{AU17}
\alpha\,\,&=&\,\,\frac{1}{2}\,\left(  \kappa -1 ) \,\,\,+\,\,\,\sqrt{(1 - \kappa)^2
\,\,-\,\,4\,\omega} \right)\,;\nonumber\\
\beta\,\,&=&\,\,\frac{1}{2}\,\left( (\kappa -  1) \,\,\,-\,\,\,\sqrt{(1 - \kappa)^2
\,\,-\,\,4\,\omega}\right)\,;
\eea

The condition that gives us the spectrum of $\omega$ is
\beq \label{AU18}
 Z(\omega, 1 - u=1)\,\,\, =\,\,\,\,{}_2F_1\left(\alpha, \beta, \kappa, 1
\right)\,\,\,=\,\,\,\frac{\Gamma(\kappa)\,\,\Gamma(1)}{\Gamma(\kappa - \alpha)\,\,\Gamma(\kappa -
\beta)}\,\,=\,\,0
\eeq
The equation for the eigenvalues has the form
\beq \label{AU19}
\kappa\,\,-\,\,\,\alpha\,\,=\,\,-\,\,n  \,\,;\,\,
\,\,\mbox{}\,\,\,\,\,\,\mbox{which gives}\,\,\,\omega_n\,\,=\,\,-\,\lambda_n\,\,=\,\,n^2\,+n (\kappa
+1)\,+1\,\,\,\,\mbox{where}\,\,n\,=\,0,1,2,\dots
\eeq

One can see that this spectrum practically coincides with \eq{AU6}. Therefore, we can assume that
\eq{AU7} or \eq{AU6} determines the spectrum of our problem in the entire region of $u$. This fact we
will use below to develop the numerical approach to the master equation.

\subsection{Approximate method}

The semi-classical
 solution we can compare with a simpler approximation suggested in Ref. \cite{L4}.  This approach
consists of two steps. The first one is to find the asymptotic solution, namely, the solution to the 
following equation
\beq \label{AS1}
- \kappa\,\,\phi^{'}_u\,\,+\,\,(\phi^{'}_u)^2\,\,=\,\,0
\eeq

\FIGURE[h]{
\begin{minipage}{85mm}{
\centerline{\epsfig{file=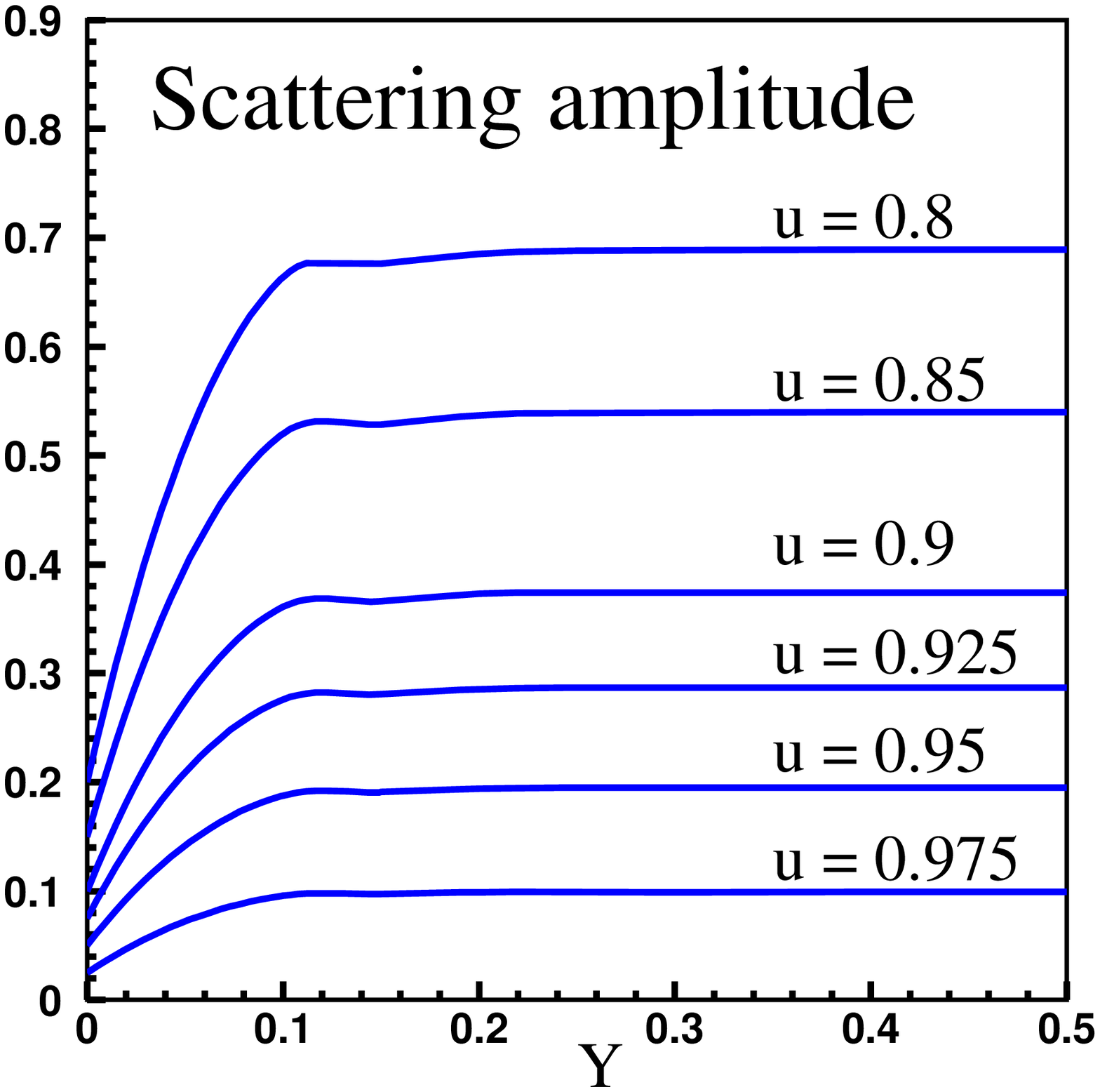,width=80mm}}
}
\end{minipage}
\caption{The scattering amplitude as a function of rapidity at different values of $u \to 1$. Recall 
that here we denote by $Y \equiv {\cal Y}$ (see \protect\eq{CALY}).  $\kappa$ is  equal to 16 in this picture.  }
\label{u1}
}

This solution is $\phi_{asymp}(u)\,\,=\,\,\kappa\,(u - 1)$ which satisfies  the normalization $\phi(u = 1)
= 0$ 
that 
follows from $Z(u=1,Y) = 1$. The next step is to find a solution in the form  $\phi(u;Y) \,\,=\,\,
\phi_{asymp}(u)\,\,+\,\,\Delta \phi(u;Y)$ assuming that $\Delta \phi(u;Y) \,\,\ll\,\,\phi_{asymp}(u)$.
The equation for  $\Delta \phi(u;Y)$ is the typical Liouville equation 
\beq \label{AS2}
\frac{\partial \,\Delta \phi(u;Y)}{\partial\,{\cal Y}}\,\,=\,\,\kappa\,\,u\, (1 - u)\,\frac{\partial 
\,\Delta 
\phi(u;Y)}{\partial\,u}
\eeq
with the initial condition
\beq \label{AS3}
\Delta \phi(u;Y = 0)\,\,=\,\,\ln( u ) \,\,-\,\kappa\,(u - 1)
\eeq
The solution is an arbitrary function of the following type 
\beq \label{AS4}
\Delta \phi(u;Y = 0)\,\,=\,\,H\left( Y\,\,+ \,\,\frac{1}{\kappa}\,\ln\left(\frac{u}{1 - u} 
\right)\right)
\eeq
which can be found from the initial condition of \eq{AS3}.

The solution has the form
\beq \label{ASZ}
Z(u;Y)\,\,=\,\,\frac{u\,e^{\kappa\,{\cal Y}}}{1\, -\, u \,+\, u\,e^{\kappa\,{\cal
Y}}}\,\exp\left(\kappa\,u\,(1 -u)\frac{ 1 - e^{\kappa\,{\cal Y}}}{1 \,-\,u\,+\,u\,e^{\kappa\,{\cal Y}}} 
\right)
\eeq

This solution corresponds to the general solution, given by \eq{SCGRF2}, but  only one solution 
$\phi^+(\omega,u)$ is taken into account. In spite of the fact that the full semi-classical solution looks 
quite different from the simple expression of \eq{ASZ} the both solution are in less than 1\% agreement at
least for ${\cal Y}\,\geq\,0.3$.

\section{The exact  solution to the problem}

A master equation of \eq{GASL} we can re-write in more convenient form, introducing a new function
${\cal G}$, namely \footnote{One can see in Appendix C that \eq{GS1} is a particular choice of the 
general transformation.},
\bea \label{GS1}
Z\,\left(\omega, u \right)\,\,\,&=&\,\,e^{ \frac{\kappa}{2}\,u}\,{\cal G}(\upsilon,\omega )\,\,u (1- 
u)\,\,\nonumber \\
&
=&\,\,\frac{1 - \upsilon^2}{4}\,\,{\cal G}(\upsilon,\omega )\,\,e^{ \frac{\kappa}{4}\,(1 - 
\upsilon)} \nonumber \\
\mbox{where} \,\,& &\,\,\upsilon\,\,=\,\,\,\,1\,\,-\,\,2\,u
\eea

~

For the  function ${\cal G}$ we have
\beq \label{GS2}
\omega\,{\cal G}(\omega, \upsilon)\,\,=\,\,-\,\frac{\kappa^2}{16}\,( 
1\,\,-\,\,\upsilon^2\,)\,{\cal G}(\omega, \upsilon)\,\,+\,\,\left((1\,\,-\,\,\upsilon^2\,)\,{\cal 
G}(\omega, \upsilon)\right)_{\upsilon \upsilon}
\eeq

The equation for ${\cal G}$ can be re-written in the form
\beq \label{GS3}
( 1 - \upsilon^2)\,{\cal G}_{\upsilon \upsilon}\,\,\,-\,4\,\upsilon\,{\cal G}_{\upsilon}
\,\,+\,\,\left\{ - \omega\,-\,2\,-\,\frac{\kappa^2}{16} ( 1 - \upsilon^2 ) \right\}\,{\cal G}\,\,=\,\,0
\eeq

For function $S_{n,m}\,\,=\,\,( 1 - \upsilon^2)^{\frac{m}{2}}\,\,{\cal G}(\omega,\upsilon)$
\eq{GS3} reduces to the equation for the prolate spheroidal wave function with the fixed
order parameter $m \,\,=\,\,1$ and arbitrary degree parameter $n$ which is an
integer, $S_{n,m}(c,\upsilon)$  (see
Refs. \cite{AS,SPHFUN}),
namely,
\beq \label{GS4}
\frac{d}{d \upsilon}\,\left((1 - \upsilon^2)\,\frac{d S_{n,m}(c,\upsilon) }{d
\upsilon}\right)\,\,+\,\,\left(\,\lambda^m_n\,\,-\,\, c^2 \,\upsilon^2\,\,-\,\,\frac{m^2}{1 - \upsilon^2}
\right)\,\,\,S_{n,m}(c,\upsilon)\,\,=\,\,0
\eeq
where
\bea \label{GS5}
\omega_n\,\,&=&\,\,-\,\frac{\kappa^2}{16} \,\,-\,\,2\,\,+\,\,m(m +
1)\,\,-\,\,\lambda^m_n\,\,=\,\,\,\,-\,\frac{\kappa^2}{16}\,\,-\,\,\lambda^m_n\,\,; \nonumber
\\
m\,\,&=&\,\,1\,\,;\,\,\,\,\,\,\,\,\,\,\,\,\,\,\,\,c^2\,\,\,=\,\,\,-\,\,\frac{\kappa^2}{16}\,\,.
\eea
To verify \eq{GS5} we can substitute  $S_{n,m}\,\,=\,\,( 1 - \upsilon^2)^{\frac{m}{2}}\,\,
{\cal G}(\omega,\upsilon)$ into \eq{GS4} and obtain the following equation for fuinction ${\cal G}(\omega,\upsilon)$
\beq \label{GS51}
( 1 - \upsilon^2)\,{\cal G}_{\upsilon \upsilon}\,\,\,-\,2\,(m + 1)\,\upsilon\,{\cal G}_{\upsilon}
\,\,+\,\,\left\{ \lambda^m_n\,\,-\,\,c^2\,\upsilon^2\,-\,m( m + 1) \right\}\,{\cal G}\,\,=\,\,0
\eeq
From \eq{GS4} and \eq{GS5}
the set of the eigenfunctions for the generating function $Z_n$ is equal to
\beq \label{GS6}
Z_n\,\,\,=\,\,( 1 -
\upsilon^2)^{\frac{1}{2}}\,\,\,S_{n,1}(i\frac{\kappa}{4},\upsilon)\,\,\,\,e^{\frac{\kappa}{4}\,(1 -
\upsilon)}
\eeq

The normalization of function $S_{n,1}$ is given by the  following equations \cite{AS}
\beq \label{GS8}
\|S_{n,1}(c,\upsilon)\|^2\,\,\,=\,\,\,\int^1_{-1}\,\,\,d\,\upsilon\,\,|S_{n,1}(c,\upsilon)|^2\,\,\,\,\,=
\,\,\,\frac{2\,n\,(n + 1)}{2 n +1}
\eeq

The Green function of \eq{GS4} has the form
\beq \label{GS9}
G\left(Y - Y';\upsilon, \xi \right)\,,\,=\,\,\,\sum^{\infty}_{n=1}\,e^{ - \lambda^1_n \,\left( Y - Y'
\right)}\,\,\frac{S_{n, 1} \left(i\frac{\kappa}{4},\upsilon \right)\,\,\,S_{n, 1}
\left(i\frac{\kappa}{4},\xi \right)}{
\|S_{n,1}(i\frac{\kappa}{4},\upsilon)\|^2}
\eeq

It should be stressed that the $Y$ dependence of the Green function is determined by the condition  $E_n -
E_{min} = - \omega_n + \omega_0 = \lambda^1_n $ in  accordance with the Schroedinger equation
(see \eq{SCHEQ}). $E_0 = -\omega_0$ is the  lowest energy level which corresponds to the asymptotic
state of \eq{ZASP}. Therefore, \eq{GS9} explicitly demonstrate that the generating function $Z$
tends to $Z \to Z_{asymp}$ at high energies.

This Green function leads to the following formula for the generating function $Z(Y,u)$
\beq \label{GS10}
Z\left(Y, u \right)\,\,\,= \,\,\,\,\,u\,\,\,+\,\,\,
\eeq
$$
\,\,\frac{u (1 - u)}{\sqrt{1 - ( 1 - 2u)^2}}\,\,e^{\frac{\kappa}{2}\,\,u}\,\,
\sum^{\infty}_{n=1}\,\left(\frac{1 \,\,-\,\,e^{ - \lambda^1_n\, Y
}}{\lambda^1_n}\right)\,\,\frac{S_{n,
1}
\left(i\frac{\kappa}{4},1 - 2 u \right)\,\,\int^1_{-1}\,\,d\,\xi\,\,S_{n, 1}
\left(i\frac{\kappa}{4},\xi
\right)\,\left(- \,\kappa\,\sqrt{1 - \xi^2}\,\,e^{-\,\,\frac{\kappa}{4}\,\,(1 -
\xi)}
\,\right)}{\|S_{n,1}(i\frac{\kappa}{4},1 - 2 u)\|^2}
$$

Fortunately, we have a rather detailed knowledge of the spheroidal functions in mathematics including
the spectrum of $\lambda^1_n$ (see Refs. \cite{AS,SPHFUN}). The series in \eq{GS10} is well converged
and, practically, we need
to take into account only about 10 terms.

\eq{GS10} can be easily generalized for arbitrary  initial condition: $Z(Y = 0; u)\,\,=\,\,F(u)$.

Indeed, the solution has the form
\beq \label{GS11}
Z\left(Y, u \right)\,\,\,= \,\,\,\,\,F(u)\,\,\,+\,\,\,\,\,\frac{u (1 - u)}{\sqrt{1 - ( 1 -
2u)^2}}\,\times
\eeq
$$
\times\,\,e^{\frac{\kappa}{2}\,\,u}\,\,
\sum^{\infty}_{n=1}\,\left(\frac{1 \,\,-\,\,e^{ - \lambda^1_n\, Y
}}{\lambda^1_n}\right)\,\,\frac{S_{n,
1}
\left(i\frac{\kappa}{4},1 - 2 u \right)\,\,\int^1_{-1}\,\,d\,\xi\,\,S_{n, 1}
\left(i\frac{\kappa}{4},\xi
\right)\,\left(\,\sqrt{1 - \xi^2}\,L\left(F(\frac{1 - \xi}{2})\right)\,e^{\,\,\frac{\kappa}{4}\,\,(1 -
\xi)}
\,\right)}{\|S_{n,1}(i\frac{\kappa}{4}, 1 - 2 u)\|^2}
$$
where $L$ is the Sturm - Liouville operator defined in \eq{GF4}. In coming papers we will discuss
the master equation with different initial conditions that reflect the interaction with nuclei and
other targets \cite{KL1,KL2}

\section{Numerical estimates}
In our numerical calculations we use the package for mathematica written by P. Falloon
(see Ref. \cite{SPHFUN})  which allows us to compute both the eigenvalues $\lambda^1_n$ and the
spheroidal wave functions $S_{n,1}$ with fixed $c $ given by \eq{GS5}.
The result of our calculations at fixed $\kappa = 16$ which is related to the QCD coupling constant
$\alpha_s = 0.25$, are plotted in \fig{sph1} and \fig{sph2}.

\FIGURE{\begin{tabular}{c c}
\epsfig{file=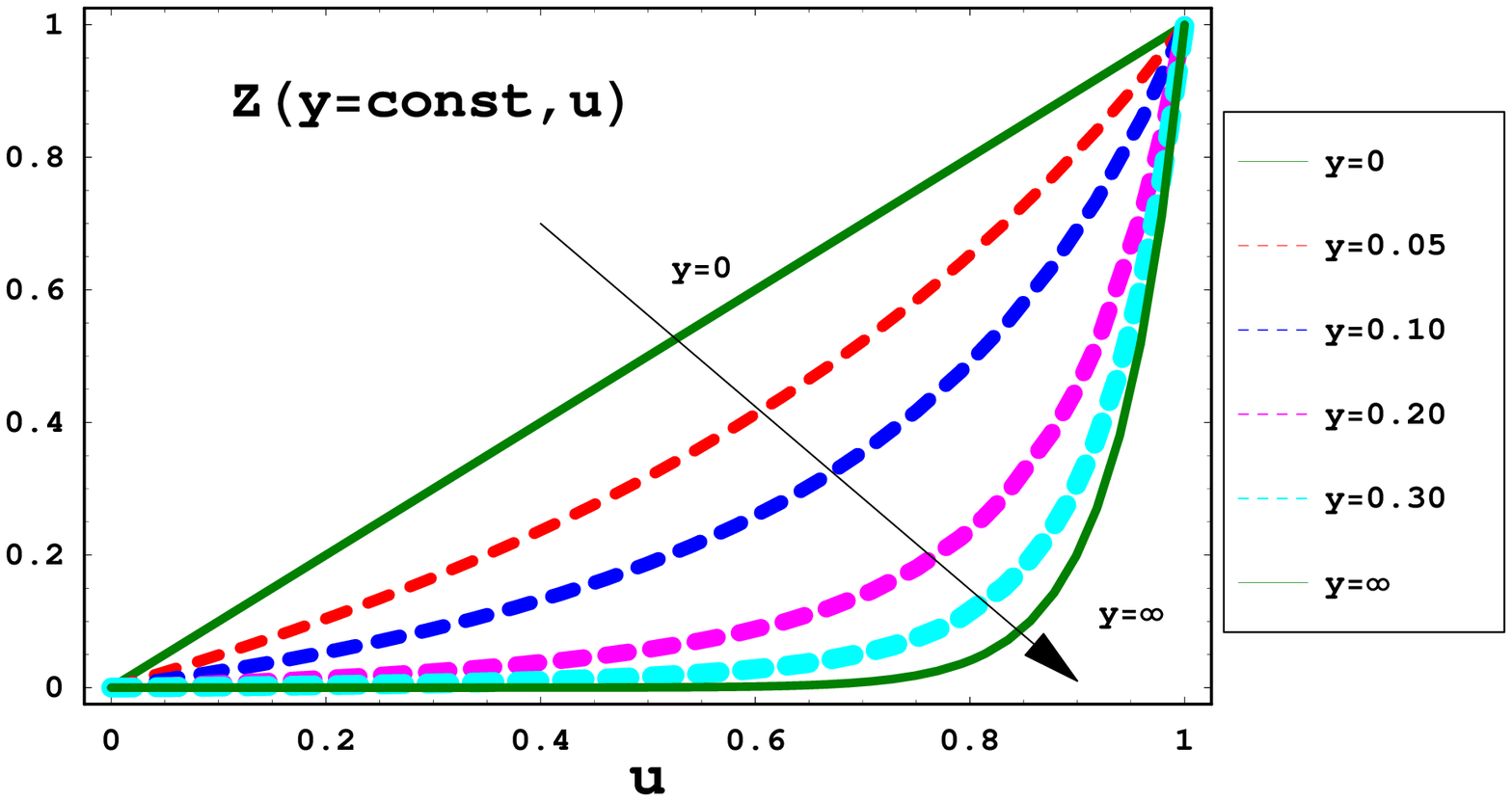,width=85mm} & \epsfig{file=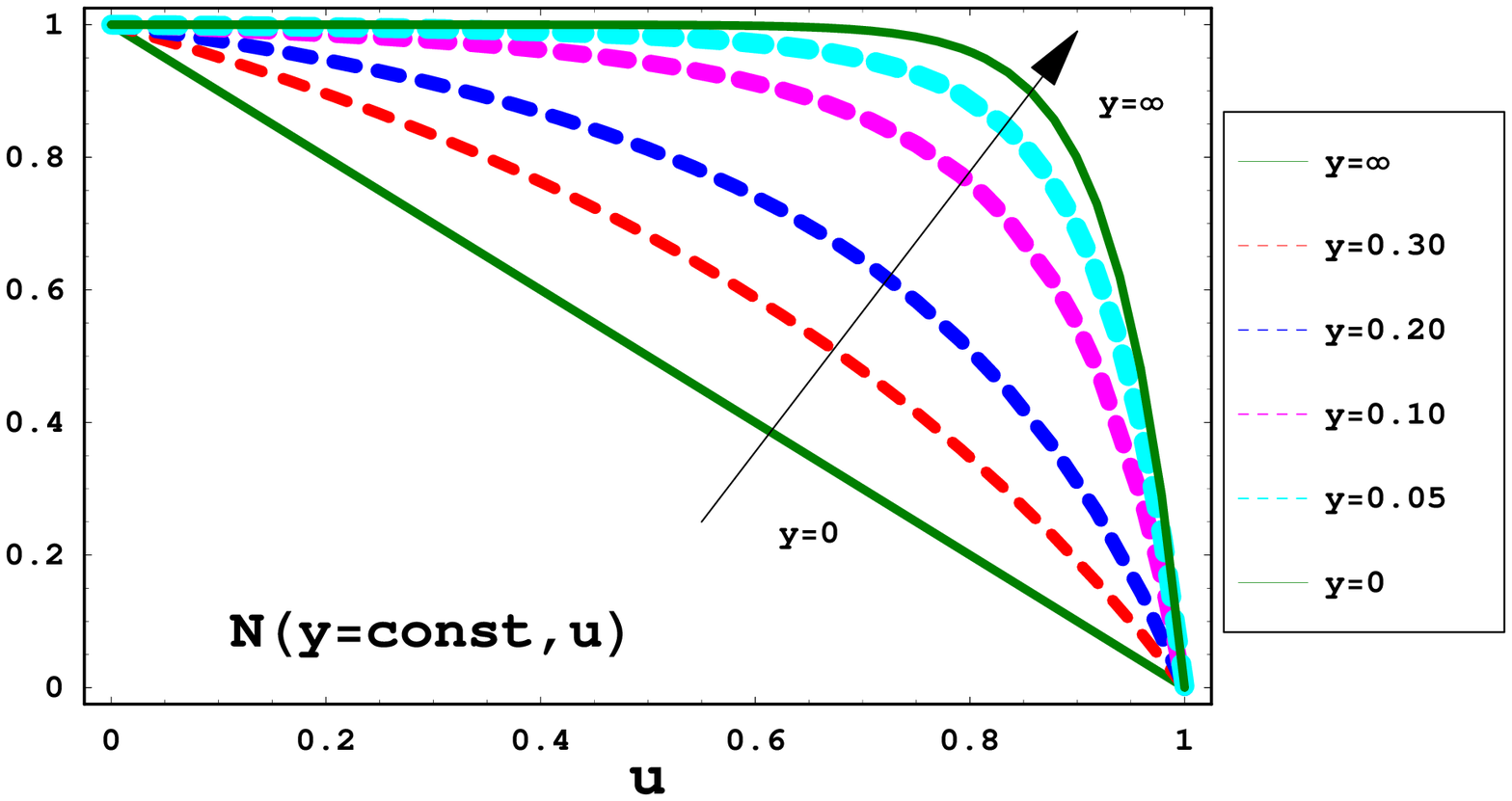,width=85mm} \\
\fig{sph1}-a & \fig{sph1}-b \\
\end{tabular}
\caption{ The $u$ dependence for the generating function $Z(Y,u)$ (\fig{sph1}-a) and the scattering
amplitude $N(Y,u) = 1 - Z(Y,u)$ (\fig{sph1}-b) at different values of rapidity. In this figure we
plot $Y \equiv
{\cal Y}$  (see \protect\eq{CALY} ). At $Y = \infty$ the asymptotic solution of \protect\eq{ZASP} is
shown.$\kappa$ = 16.}
\label{sph1}
}

\fig{sph1} shows the generating function as function of $u$ at different energies. One can see that
at ${\cal Y} \,\gg\,1$ the solution approaches the asymptotic solution of \eq{ZASP} ( the low
dashed
line in \fig{sph1}) in accordance with our expectations \cite{AMCP,BOR,L4}. To understand the scale
of the energy dependence we need to go back to the rapidity $Y$ from ${\cal Y}$. As we have
discussed
\beq \label{YY}
Y\,\,\,=\,\,\,\frac{\kappa}{\Gamma( 1 \to 2)}\,{\cal
Y}\,\,\,\,=\,\,\,\frac{1}{\alpha^2_S\,\omega_{BFKL}}\,\,{\cal Y}
\eeq
where $\omega_{BFKL}$ is the intercept of the BFKL Pomeron\cite{BFKL}, namely,
$\omega_{BFKL}\,\,\,=\,\,4\,\ln 2\,\bas$. For $\alpha_S = 0.25$ which we use in this paper,
$Y\,\,=\,\,23\,{\cal Y}$. Therefore, $ {\cal Y} = 0.3$ gives $Y - Y_0 \,=\, 6.9$ ($Y_0$ is the
initial rapidity from which we can apply our high energy approximation).

\fig{sph2} shows the main result of this paper: the energy (rapidity) dependence of the scattering
amplitude. One can see that the exact solution approaches the asymptotic solution as has been
foreseen in the approximate solutions to the problem (see, for example, Refs. \cite{AMCP,BOR,L4}).

This result  eliminates the last  uncertainty
 that the high energy limit of the scattering amplitude does not show the black disc
behaviour but we have rather gray disc one as has been discussed before in Ref. \cite{L4}.
It turns out that at ${\cal Y} \,\geq\,0.5$ ($Y \geq\,12.5$) the difference between the exact
solution and the asymptotic one is very small. However, for ${\cal Y} \,\leq\,0.5$ the difference
between the semi-classical approach and the exact solution is rather large and all approximate
methods should be taken with the justified  suspicion in this kinematic region.

\FIGURE{
\centerline{\epsfig{file=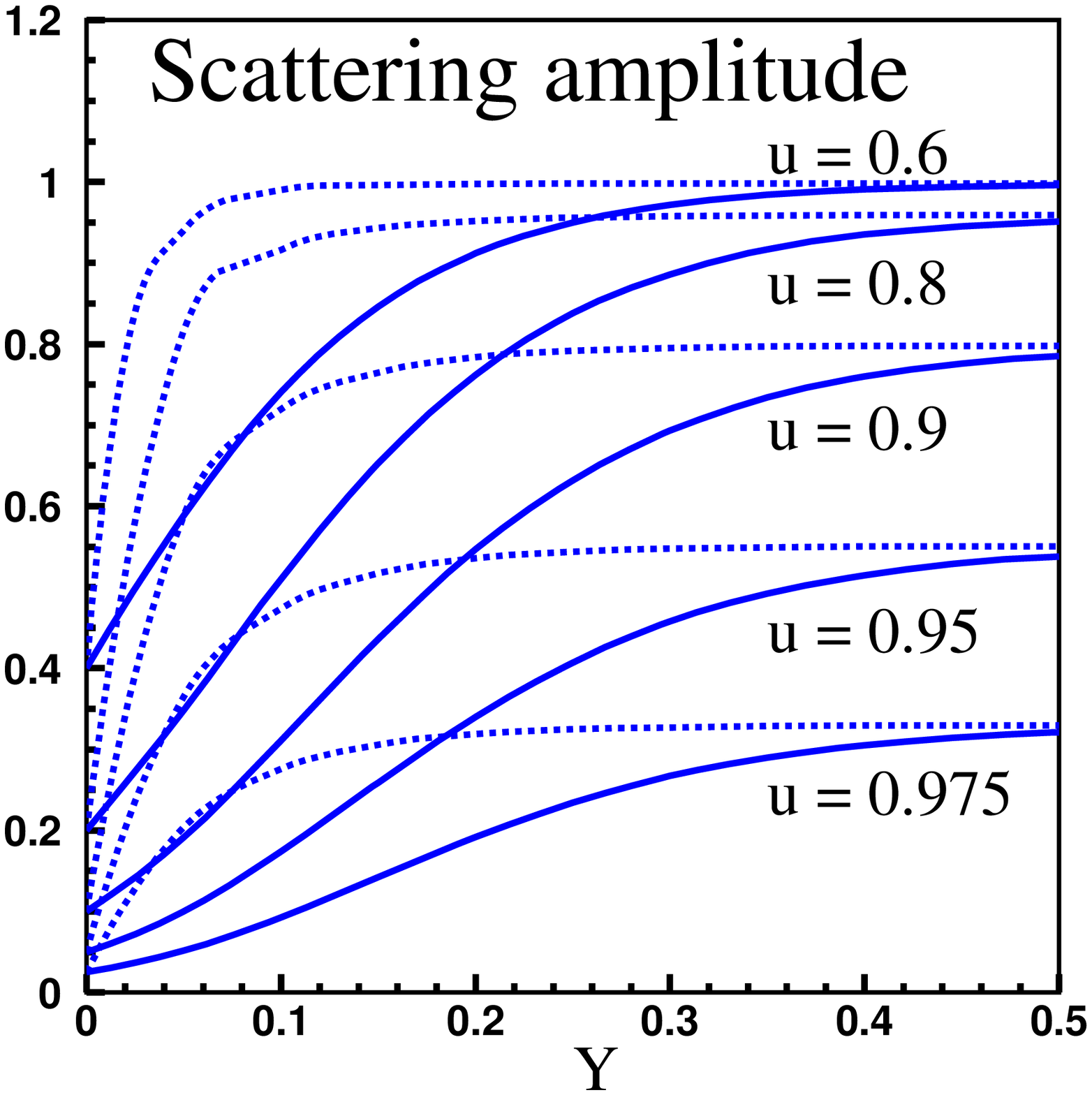,width=160mm}}
\caption{ The rapidity dependence of the scattering amplitude $N(Y,u)$ at different values of $u$.
The dashed curves are the semi-classical solution of \protect\eq{SCGRF2}, while the solid lines are
the exact solution to the master equation.  In this figure we plot $Y \equiv
{\cal Y}$  (see \protect\eq{CALY} ) and $\kappa$ = 16.  }
\label{sph2}
}

In Appendix D we  develop the numerical method specially for small values of $Y$ and we compare it
with the exact solution of
\eq{GS10}.
\section{Comparison with approximate approaches}
\subsection{Exact solution and the semi-classical approach}
As one can see from \fig{sph2} the semi-classical approach preserves the singularity structure of the exact solution and correctly reproduces the high energy limit of the amplitude at any value
of $u$  but it cannot describe the approaching to this limit. The reason for such a behaviour is clear if we compare the spectrum
$\lambda_n$ at
low $n$ of the semi-classical approach given by \eq{SOM} with the values of $\lambda_n$ of the exact solution (see Table 1). In the
semi-classical solution at small values of $n$\,\, $\Delta \lambda_n\,=\,\lambda_{n +1} - \lambda_n = \kappa$ while in the exact
spectrum $\Delta \lambda_n\,<\,\kappa$.

\subsection{Exact solution and the mean field approximation (MFA)}

For the mean field approximation we need to neglect in \eq{GA1} the second term in the r.h.s. of the equattermThe solution is very
simple (see Refs.\cite{MUCD,L1}), namely
\beq \label{COM1}
Z\left(Y; u \right)\,\,\,=\,\,\,\frac{u\,e^{ - \kappa\,Y}}{ 1\,\,+\,\,u\,\left( e^{ - \kappa\,Y}\,\,-\,\,1 \right)}
\eeq
In \fig{sph3} we plot the exact solution (solid lines) and the mean field approximation (see \eq{COM1}).

\FIGURE{
\begin{tabular}{c c}
\epsfig{file=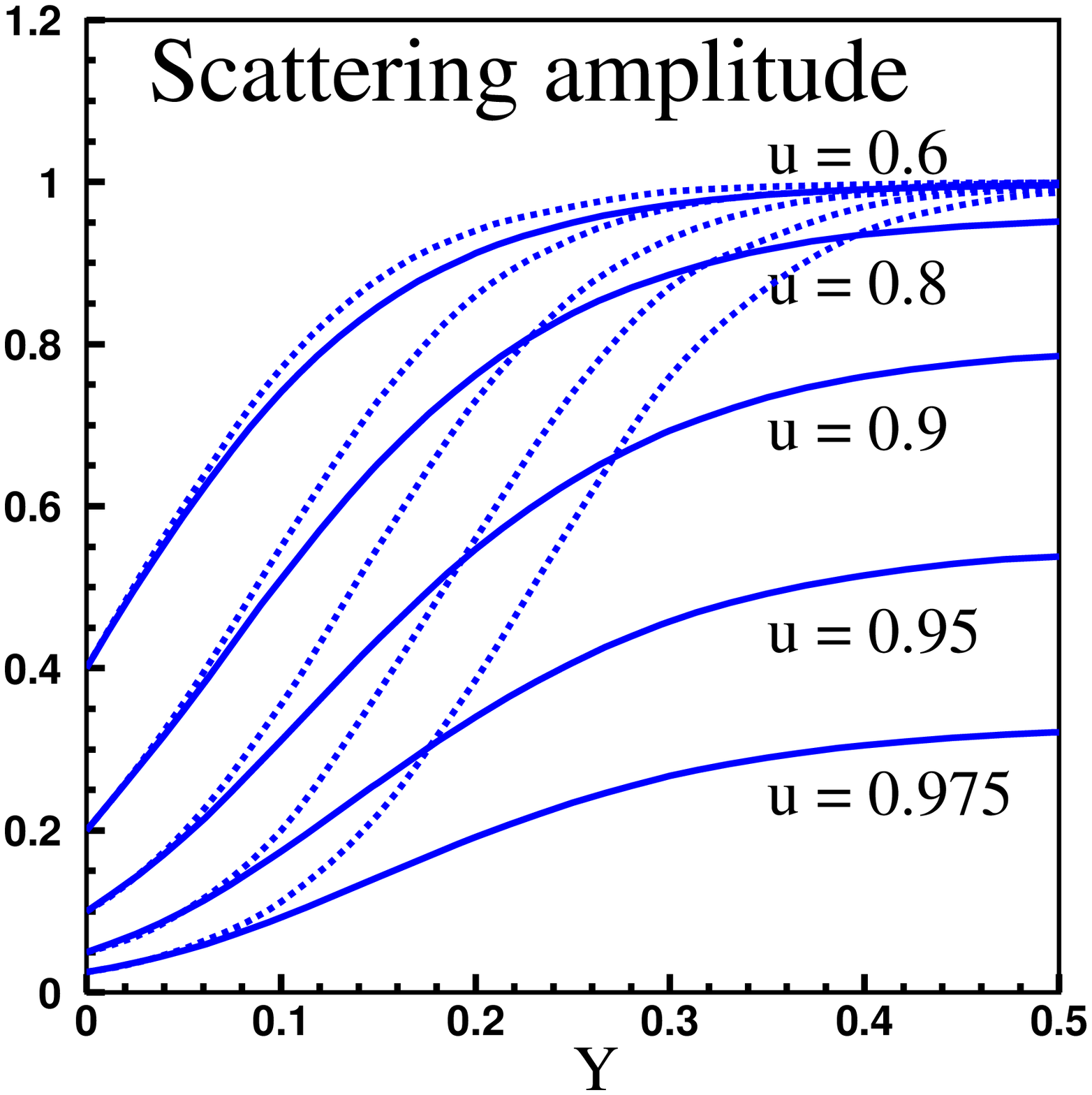,width=80mm} & \epsfig{file=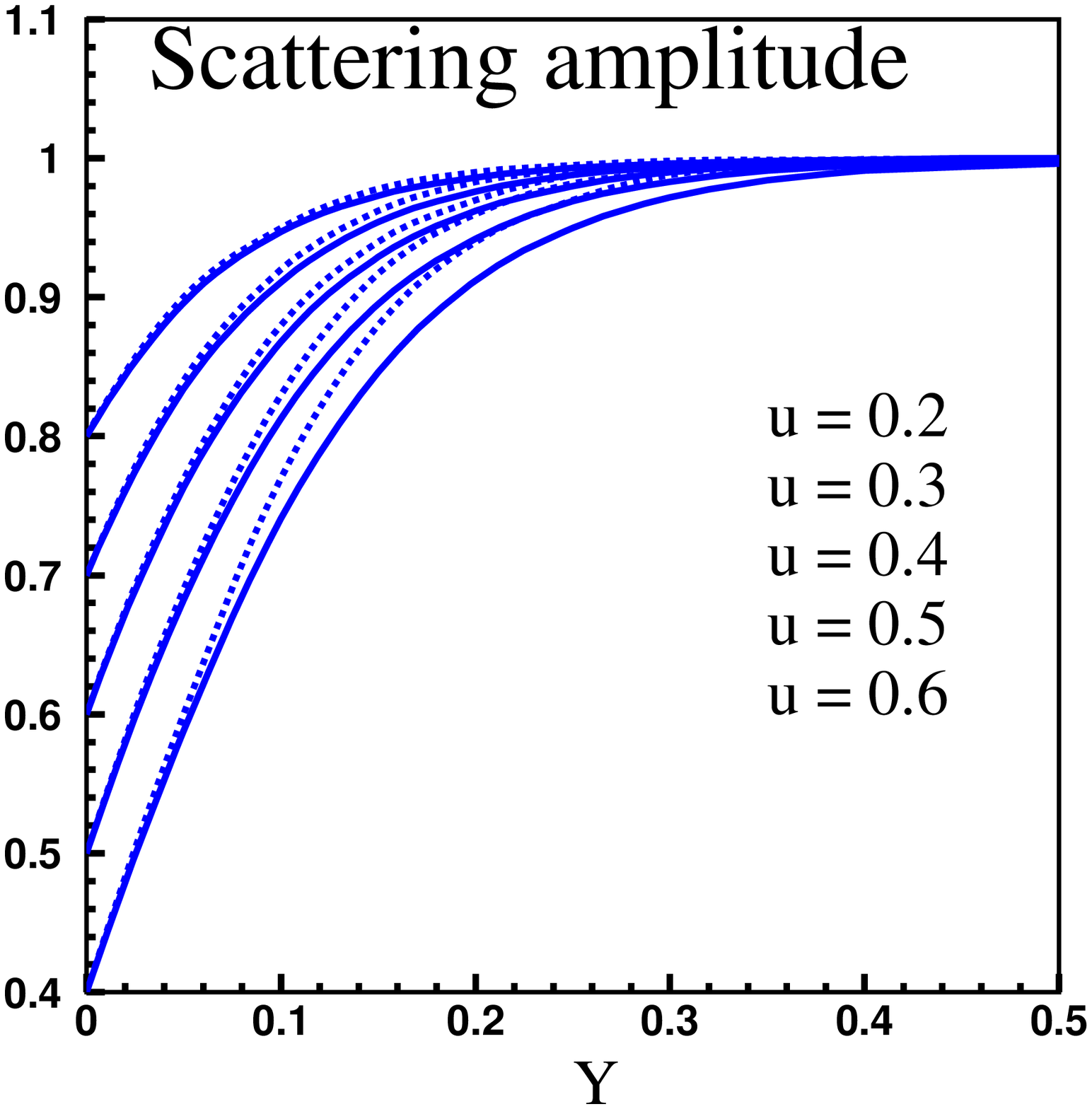,width=80mm}
\end{tabular}
\caption{ The rapidity dependence of the scattering amplitude $N(Y,u)$ at different values of $u$.
The dashed curves are the mean field approximation given by  of \protect\eq{COM1}, while the solid lines are
the exact solution to the master equation.In this figure we plot $Y \equiv
{\cal Y}$  (see \protect\eq{CALY} ) and $\kappa$ = 16.  }
\label{sph3}
}
 
One can see that the mean field approximation cannot describe the scattering  amplitude for values of $u$
 close to 1,  but it
gives a good approximation in the region of $u \,\leq\,0.6$ in the wide energy range. However, even
for large values of $u$ the MFA fails to reproduce a correct approaching the scattering amplitude to
the asymptotic limit.

The physical reason for such a behaviour is clear. The typical value for the scattering amplitude at low energy $\gamma = 1 -u
\,\approx 1/\kappa\,\,\leq\,\,1$ for the dipole-dipole scattering. Therefore for the dipole-dipole scattering the exact solution
leads to a different behaviour than the mean field approximation. However, if we have the interaction of dipole but with the nucleus
target the amplitude at low energies is proportional to $A^{\frac{1}{3}}\,\gamma(dipole-dipole)\,= \,A^{\frac{1}{3}}/\kappa\,=\,1 -
u\,\approx 1$ (see for
example the paper of Kovchegov in  Ref. \cite{BK}).  The second case when the mean field approximation can work is the deep inelastic
scattering. Indeed, we can model this process assuming that $\kappa\,=\,\,1/\alpha^2_S(Q^2_s)$. However, the scattering amplitude for
low energies is proportional to $\alpha^2_S(1/R^2)$ where $R$ is the target size. Therefore, for the running QCD coupling we can
assume that $1 - u \, = \, \gamma \,= \,\, \alpha^2_S(1/R^2)\,\,\approx\,\,1\,\,\geq\,1/\kappa$. and the mean field approximation can
describe the deep inelastic processes \cite{GLR}.
\subsection{Exact solution and the Mueller-Patel-Salam -Iancu (MPSI) approximation.}
It was shown by Mueller and Patel\cite{MUPA} ( see also Refs. \cite{MS,IM, KOLE,KOVMP}) that in a restricted region of energies we
can consider the Pomeron loops larger than $Y/2$. Such loops we can sum and the answer for our model in ouir notation  looks as
follows (see Ref.
\cite{KOVMP})
\beq \label{COM2}
N(Y,u)\,\,=\,\,1 - \frac{1}{(1 -u)\,e^{ \kappa Y}}\,\exp \left(\frac{1}{(1 -u)\,e^{ \kappa Y}}\right)\,\Gamma\left(0,\frac{1}{(1
-u)\,e^{ \kappa Y}}\right)
\eeq
\fig{sph4} shows the comparison of the solution given by \eq{COM2} (dotted lines) with the exact solution (solid lines). One can see
that only in the very limited range of energies $Y\,\leq\,0.1$ and at small values of $u$ the MPSI approximation can describe the
scattering amplitude.
\FIGURE{
\centerline{\epsfig{file=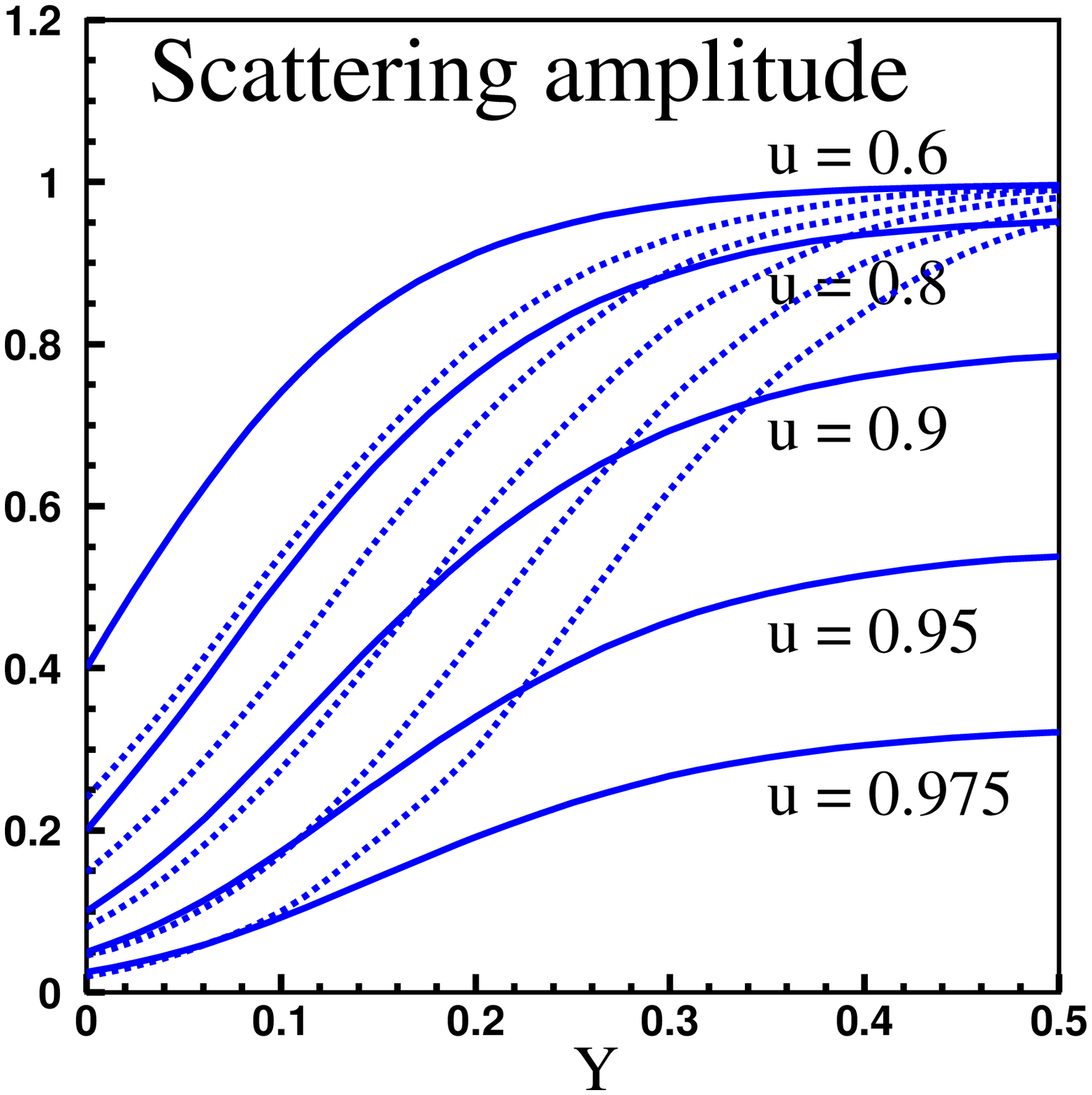,width=120mm}}
\caption{ The rapidity dependence of the scattering amplitude $N(Y,u)$ at different values of $u$.
The dashed curves are the MPSI approximation  given by  of \protect\eq{COM2}, while the solid lines are
the exact solution to the master equation.In this figure we plot $Y \equiv
{\cal Y}$  (see \protect\eq{CALY}) and $\kappa$ = 16.  }
\label{sph4}
}

\subsection{Exact solution and the  gray disc behaviour.}
One of the most interesting results of this paper is the asymptotic gray disc behaviour at ultra high energies: the solution does not go to 1 but to a constant which is smaller than 1 and which depends on the  initial conditions or, in other words, on the value of the scattering amplitude at low energies. Such a solution does not contradict to the unitarity constraints and  it is consistent with some of the previous approaches (see Refs. \cite{AMCP,L4}).
However, among the approximate approaches to the solution of the master equation situations is quite different:
most of them: the mean field approximation, the MPSI solution as well as the solution of Ref. \cite{SHXI} lead to the black disc behaviour and only the semi-classical approach gives a constant limit at high energy that is less than 1.
Therefore it is worthwhile to discuss the gray disc limit in more details.

 The scattering amplitude is equal to \cite{BK,L2}
 \bea \label{GD1}
 N\Lb Y, \gamma(Y_0) \Rb\,\,&=&\,\,1\,\,-\,\,Z\Lb Y, 1 -  \gamma(Y_0) \Rb\,\,\,=\,\,\sum^{\infty}_{k = 1}\,\,  (-1)^k\,\gamma^k(Y_0) \Lb \sum^{\infty}_{n = 1}\,\,P_n(Y)\,\frac{n!}{k! ( n -k)!}\Rb \,\,\nonumber\\
 &\stackrel{\rm \gamma(Y_0) \ll 1}{\Longrightarrow}&\,\,\gamma(Y_0)\sum^{\infty}_{n=1}\,n\,\,P_n(Y)\,\,\,=\,\,\,\,\langle |n |\rangle \times \gamma(Y_0)
\eea
where $\gamma(Y_0)$ is the scattering amplitude of one dipole with the target at low energy. If this amplitude is very small \eq{GD1} shows that the scattering amplitude for the dipole - target scattering at high energy is equal to the product of the low enegy amplitude by the average nuenergy($ \langle |n |\rangle$)  of dipoles at low energy. Therefore, to obtain the amplitude that equals to 1 at high energy we need to assume that $\langle |n |\rangle$
becomes large (at least of the order of  $1/\gamma(Y_0)$ ).  Large number of low energy dipoles is natural to expect
in the mean field approximation where the tree (`fan') diagrams contribute (see \fig{pom}). On the other hand the
mean field approximation violates the $t$-channel unitarity   or, in other words, it gives  an  incorrect normalization of the
parton wave function (see Ref. \cite{MUSH}). Therefore,  in  the correct answer we rather expect that the average number of
partions should be of the order of 1. Indeed,  the enchanced diagrams of \fig{pom} are the most essential
at high energies to protect the   $t$ - channel unitarity \cite{GRC}. In these diagrams we expect the average number of dipoles at low energy to be the same as in our initial condition, namely, $\langle |n |\rangle \approx 1$.
In this situation the behaviour of the amplitude at high energies depends on the value of the scattering amplitude at low energies as we see in our exact solution.

\EPSFIGURE[ht]{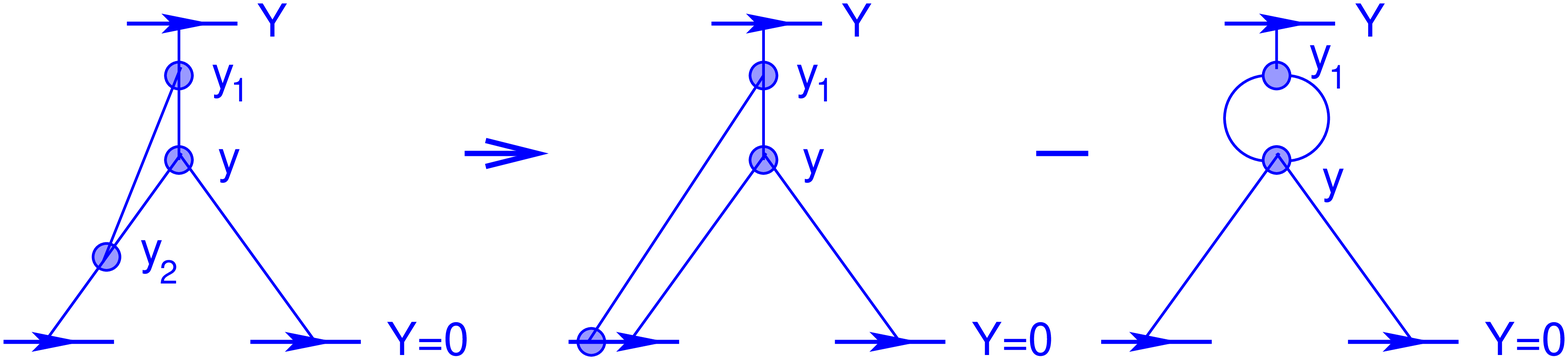, width=140mm}{The first correction to the $ P \to 2P$ vertex due to the merging of two Pomerons into one Pomeron. Solid lines are Pomerons. Lines with arrows denote the interacting dipoles.\label{3pcor}}

This result we could also foresee if we considered the first corrections to the vertex of splitting of one Pomeron to two Pomerons (transition $P \to 2P$, see \fig{3pcor}).  First thing that we can see from this picture,  which graphically shows the result of integration with respect to $y_2$,  that the first corrections  lead to an increase of the value of
the $2 P \to P$ vertices  with rapidities $y_1$ ( the first diagram in \fig{3pcor}) and $y$ ( the second diagram in  \fig{3pcor}). Indeed, the simple diagram in \fig{3pcor} has the following explicit expression
\bea \label{GD12}
A\Lb \fig{3pcor} \Rb\,\,&=&\,\,\Gamma^2(1 \to 2)\,\Gamma(2 \to 1)\,\int^Y_y\,\,d\,y_1\,\int^{y}_0\,\,d y_2\,\,e^{\Gamma(1 \to 2)
\Lb Y + y_1 + y - y_2\Rb} \,\,\nonumber \\
 &=&\,\,\Gamma(1 \to 2)\,\Gamma(2 \to 1)\,\int^Y_y\,\,d\,y_1\,\left\{e^{\Gamma(1 \to 2)\,\Lb
Y + y_1 \Rb}\,\,-\,\,e^{\Gamma(1 \to 2)\,\Lb
Y + y_1  + y\Rb}\,\right\}
\eea

 Having this observation in mind  we solve the mean field equations but assuming that the vertex for
$P \to P$ transition ($\Delta$) is not equal to the vertex of splitting of one Pomeron to two Pomerons $\Gamma(1 \to 2)$. The equation has the form
\beq \label{GD2}
\frac{\partial Z\Lb Y;u \Rb}{\partial Y}\,\,\,=\,\,\Lb - \Delta\,u \,\,+\,\,\Gamma(1 \to 2)\,u^2 \Rb \,\,\frac{\partial Z\Lb Y;u \Rb}{\partial u}
\eeq
\eq{GD2} has an obvious solution, which for the scattering amplitude has the form
\beq \label{GD3}
 N\Lb Y, \gamma(Y_0) \Rb\,\,\,=\,\,\,\frac{\gamma(Y_0)\,e^{\Delta\,\,(Y - Y_0)}}{1\,\,+\,\,\,
 \gamma(Y_0)\,\,\Gamma(1 \to 2)\,\,
 \Lb e^{\Delta\, \,(Y - Y_0)}\,\,-\,\,1\,\Rb}
\stackrel{\rm  Y- Y_0 \,\gg\,1}{\longrightarrow}\,\,\,\frac{\Delta}{\Gamma(1 \to 2)}
\eeq
One can see that if $\Gamma(1 \to 2) > \Delta$ we expect a gray disc. It is easy to check that the MPSI approximation with
different vertices for $P \to P$ and $ P \to 2P$ transitions give the same answer as \eq{GD3}.
Therefore, \eq{GD12} shows that the gray disc behaviour is very likely for the exact solution.

Formally speaking we expect to have two asymptotic solutions  ($Z_{\infty}$)  which stem from the master equation with $\partial Z_{\infty}/\partial Y$=0, since the equation is the second order differential equation.   The linear combination
which satisfy the boundary conditions $Z_{\infty}( u=0)=0$ and $Z_{\infty}( u=1)=1$, is the solution to the problem at high energy, which leads to a gray disc behaviour of the scattering amplitude. These two boundary  conditions are very natural since they say: (i) for $u=0$ , that if amplitude is equal to 1 at $Y=0$ none of interactions can change this fact due to the $s$-channel unitarity; and (ii) for $u=1$, that starting with zero amplitude for dipole interaction at $Y=0$, we cannot obtain an amplitude different from zero in an interacting system. Therefore, neither the fact of two
 asymptotic solutions  nor the boundary conditions depend on the particular form of interaction. For example, they do not depend on whether we took into account the possibility for two Pomerons to re-scatter in two Pomerons as it is
done in this paper, or whether we neglect such interaction as was assumed in Ref. \cite{SHXI}.
In both cases we have the asymptotic solution, that satisfies the boundary conditionssatisfiesform of the gray disc.
One of the main result of this paper is the proof that nothing unusual happens with the system in the process of evolution from $Y=0$ to very large values of $Y$ and the resulting solution approaches the asymptotic one. To understand this result it is instructive to share with our reader what kind of an alternative solution we could expect.  Frankly speaking, we can expect in such kind of problem the asymptotic solution that has a singularity in the  singular points of this equation ($u =0$ and $u =1$). Even more, the solution of the mean field approximation, for which we have the same boundary condition, has an asymptotic solution, which we can write in the form
\beq \label{GD4}
Z_{\infty}(u) \,\,=\,\,\Theta( 1 - u)
\eeq
One can see that $Z_{\infty}(u) = 0 $ for $u <1$ and $Z_{\infty}(u) =1$ for $u =1$.  However, this solution has an singular behaviour and $Z'_u = \delta(1 - u)$ . Nevertheless, since $(1 - u)\,\delta(1 - u) = 0$ this is a solution. The solution that we found shows that in general case there is no such singular solution. Of course, the last claim is based on our believe that the physical problem has  one and the only one solution. One can check that $(1 - u)(d^2\Theta( 1 - u)/d^2 u)\,\to\,i\infty$ at $ u \to 1$.

Our conclusion is very simple: the asymptotic behaviour of the exact solution is not only the gray disc but
this kind of behaviour gives us a check how well an approximate method is able to discuss the high energy scattering.
However, it should be very carefully stated that the real QCD case can be different; and we have arguments for both scenarios:
black disc behaviour \cite{FR,KAN} and the gray disc one \cite{L4}.  The difference  is in the extra integration over sizes of produced dipoles. The scattering amplitude in the Born approximation of perturbative QCD is very large for the dipoles of the large sizes. Therefore,  the equation of \eq{GD1} - type can give the black disc even if the average number of dipoles is about 1 since it can be one dipole but of very large size. However, the question, whether we can trust the perturbative QCD calculations for such dipoles , is the open question.

\section{Conclusions}

The core of this paper is the exact analytical solution to the BFKL Pomeron calculus in the zero
transverse dimension given in section 4.  The method, that we proposed, is very general and can be
applied to a various  kind of  practical problems in future.

The scattering amplitude at high energies, which stems from  the exact solution, shows the
gray disc behaviour  of the asymptotic solution (see \eq{ZASP}). Therefore, we filled the hole in
the
proof demonstrating that the asymptotic behaviour of \eq{ZASP} satisfies the initial conditions.

We found thainitialapproximations to the master equation,  developed in the past (see Refs.
\cite{AMCP,BOR,L4,REST,SHXI}),  work quite well at large values of rapidity ($Y\,\geq\,10$) and can
be used in
either  asymptotic estimates or for an illustration of the main properties of the high energy
interaction. The advantage of all of them is their simplicity and transparent physics ideas that are
behind these methods. In this respect the semi-classical approach is  especially attractive since
it gives the clear understanding of the spectrum of the problem and the behaviour of the
eigenfunctions.  However, the comparison with the exact solution shows that we cannot trust the
approximate methods for low energies, namely, for $Y \,\leq \,10$ where there exist  the most of
the
experimental data including ones that we anticipate from the LHC.

We firmly believe that the solution for the BFKL Pomeron calculus at zero transverse dimension is
the necessary step in our understanding of the behaviour of the scattering amplitude at high
energies in QCD. The real QCD problem is much more complicated but the fact, that we can compare the
approximation methods with the exact solution for our toy model and found how they work,
might open a way to solve the QCD problem at high energies developing  different  approximations
(see  Refs.\cite{MUPA,IM,KOLE,L4} where one can see the first attempts of such kind of approaches).

\section*{Acknowledgments:}
We want to thank Asher Gotsman, Edmond Iancu,  Volodya Khachtryan, Uri Maor  and
 Jeremy Miller for very useful
discussions on the subject
of this paper.  The special thanks go to  Bo-Wen Xiao for discussion with us the results of Ref.
\cite{SHXI}.

This research was supported in part  by the Israel Science Foundation,
founded by the Israeli Academy of Science and Humanities and by BSF grant \# 20004019.

\appendix

\section{Appendix A: General way  for solving linear nonhomogeneous boundary problem.} 
\label{sec:A}

At this appendix we
 describe the general way for solving linear  non-homogeneous boundary value problem using the appropriate 
Green's 
functions, that can be constructed from the  eigenfunctions of the Sturm-Liouville problem. 

Here we are dealing  only with the  issues related to  solutions of the  master equation, 
which is equivalent to the BFKL Pomeron calculus in zero transverse dimension, while  for more detail 
information, related to this subject,
 one can find in Ref. \cite{POLY}.

\noindent First of all, it is easily to check this using  trivial transformations:

$$ 
s(v) \, = \, \frac{1}{f(v)} \, \exp \left[ \int \frac{g(v)}{f(v)} dv \right], \,\,\,\,\, 
p(v) \, = \, \exp \left[ \int \frac{g(v)}{f(v)} dv \right],
$$
\beq
 \,\,\,\,  q(v) \, = \, - \, \frac{h(v)}{f(v)} \, \exp \left[ \int \frac{g(v)}{f(v)} dv \right]
\eeq

\noindent the general linear non-homogeneous partial differential equation of the parabolic type:

\beq \label{A3}
\frac{\partial Z}{\partial y} \, = \, f(v) \frac{\partial^2 Z}{\partial v^2} \, + \, g(v) \frac{\partial Z}{\partial v} \, + \, h(v) Z \, + \, \varphi(y,v)
\eeq

\noindent can be reduced to the equation which is  suitable finding  a  solution to the Sturm-Liouville 
problem:

\beq \label{SLEQ1}
s(v) \frac{\partial Z}{\partial y} \, = \,\frac{\partial}{\partial v} \left[p(v) \frac{\partial Z}{\partial v} \right] \, 
- \, q(v) Z \, + \, s(v) \varphi(y,v)
\eeq

\noindent Denoting by $\Phi(y,v) = s(v) \varphi(y,v)$ we have the equation:

\beq \label{SLEQ2}
s(v) \frac{\partial Z}{\partial y} \, = 
\,\frac{\partial}{\partial v} \left[p(v) \frac{\partial Z}{\partial v} \right] \, - \, q(v) Z \, + \, \Phi(y, v)
\eeq

\noindent For arbitrary initial  $Z(0,v)=\theta(v)$ and boundary $Z(y,v_1)=\alpha,  \,\,\,\,  Z(y,v_2)=\beta$ 
conditions the general solution is given by the following expression (see Refs. \cite{KAMKE,POLY}):
\bea  \label{SLSOL1}
Z(y,v) 
\,\, &=& \int^{y}_{0} \int^{v_2}_{v_1} \Phi(\tau,\xi) \cdot G(y - \tau,  v, \xi) \, d\tau \, d\xi  \, +  \, \int^{v_2}_{v_1} 
s(\xi) \cdot \theta(\xi) \cdot G(y ,  v, \xi) \, d\xi \nonumber \\ 
&& \, + \,  p(v_1) \int^{y}_{0} \alpha \cdot \Lambda_1(y - \tau,  v) \, d\tau \, +  \, p(v_2) \int^{y}_{0} \beta \cdot
 \Lambda_2(y - \tau,  v) \, d\tau 
\eea

\noindent Here the Green's function is given by:

\beq \label{GREENF}
G(y, v, \xi) \, = \, \sum_{n=1}^{\infty} \frac{Z_{n}(v) \, Z_{n}(\xi)}{ ||Z_{n}||^2} \, 
\exp{( - \lambda_{n} y)}\,\,; \,\,\,\,\,\,\,\,\,   
  ||Z_{n}||^2\, =\, \int^{v_2}_{v_1} s(v) \cdot Z^{2}_{n}(v) \, dv \,\,;
\eeq

\noindent where the $\lambda_{n}$ and $Z_{n}(v)$ are the eigenvalues and corresponding eigenfunctions of the following 
Sturm-Liouville problem for the second order linear differential equation:
\beq
(p(v)\widetilde{Z}')' + [\lambda \, s(v) - q(v)]\widetilde{Z} = 0
\eeq
 with the boundary conditions
\beq
 \widetilde{Z}(v_1)=\alpha, \,\,\,\,  \widetilde{Z}(v_2)=\beta\,.
\eeq
 The functions $\Lambda_1$ $\Lambda_2$ are expressed via the Green's function:

\beq \label{LAMBDA}
\Lambda_{1}(y,v) \, \,\,\equiv \,\, \lim_{\xi \rightarrow v_1} 
\,\frac{\partial}{\partial \xi} G(y, v, \xi)\,; \,\,\,\,\,\,\,\,\, 
 \Lambda_{2}(y,v) \, \equiv \, \lim_{\xi \rightarrow v_2} \,\frac{\partial}{\partial \xi} G(y, v, 
\xi) \,\,.
\eeq

\noindent The most complicated part in \eq{SLSOL1}
 is the  two last terms that include functions $\Lambda_1$, $\Lambda_2$ and require
 applying of some regularization technique in calculating them.
 Fortunately, we can  find a simple transformation after which these two terms will disappear.
 Namely, subtracting from 
$Z(y,v)$ the initial conditions:
\beq \label{WZ}
 \widetilde{Z}(y,v) = Z(y,v) - Z(0,v)
\eeq

\noindent we will obviously have equation for function $\widetilde{Z}(y,v)$ with zero boundary conditions:
\beq
 \widetilde{Z}(y,v_1) = \widetilde{Z}(y,v_1) = 0
\eeq
 and zero initial conditions: 
\beq
 \widetilde{Z}(0,v) = 0 \,. 
\eeq
The only price for 
this transformation is transformation of the free term in \eq{SLEQ2}:

\beq \label{WPHI}
 \widetilde{\Phi}(y,v) \, = \, \Phi(y, v) \, + \, \frac{\partial}{\partial v} \left[p(v) \frac{\partial Z(0,v)}{\partial v} \right] \, - \, q(v) Z(0,v) 
\eeq

\noindent Thus we have  only the  first term (term which has nonzero contribution) in \eq{SLSOL1} for the function 
$\widetilde{Z}(y,v)$:

\beq \label{SLSOL2}
\widetilde{Z}(y,v) = \int^{y}_{0} \int^{v_2}_{v_1} \widetilde{\Phi}(\tau,\xi) \cdot G(y - \tau,  v, \xi) \, d\tau \, d\xi  
\eeq

\noindent and final answer obviously has the following form:

\beq \label{SLSOL3}
Z(y,v) = Z(0,v) + \widetilde{Z}(y,v) = Z(0,v) + \int^{y}_{0} \int^{v_2}_{v_1} \widetilde{\Phi}(\tau,\xi) \cdot G(y - \tau,  v, \xi) \, d\tau \, d\xi  
\eeq

\noindent Note, that using \eq{SLSOL3} we reduce Eq.~(A.3)  to the equation  for $\widetilde{Z}(y,v)$ which has the same form as 
Eq.~(A.4) but with the inhomogeneous term $ \widetilde{\Phi}(y,v)$ given by \eq{WPHI}.   The advantage of this transform is clear if we 
compare the rather compact answer of \eq{SLSOL3} with the solution in the form of  \eq{SLSOL1}.

\section{Appendix B: The exact semi-classical solution.} \label{sec:B}

At this appendix we calculate exactly functions $\phi^{\pm}(\omega,u)$ for the  semi-classical solution (see section 3.1)  and 
compare them 
with the  approximations at small and large values of $\omega$, which were used for calculation in this section.

Recall  that 
\beq \label{ROOTS}
\frac{d \, \phi^{\pm}}{d u}\,\,=\,\,\frac{\kappa}{2}\,\left\{\,\, 1\,\,\pm\,\,\sqrt{1 \,\,+\,\,\frac{4 
\,\,\omega}{\,\,\kappa^2 \,\,u\,(1 - u)\,}}\,\,\right\}
\eeq

\noindent Therefore,  functions  $\phi^{\pm}$ (see \eq{SC1} and \eq{SC4}) are  defined by the following integrals:

\beq \label{ROOTS_INT}
\phi^{\pm}(u) \,\,= \,\, - \,\, \int_{u}^{1} \frac{\kappa}{2}\,\left\{\,\, 1\,\,\pm\,\,\sqrt{1 \,\,+\,\,\frac{4 
\,\,\omega\,}{\,\, \kappa^2 \,\,u'\,(1 - u')\,}}\,\,\right\} \, du' \, = \, I_1 \, + \, I_2 
\eeq

\noindent where the first integral is trivial:

\beq 
I_1 \, = \,\, - \,\, \int_{u}^{1} \frac{\kappa}{2} \, du' \,\, = \,\, \frac{\kappa}{2} \, (u \, - \,1)
\eeq

\noindent The second term \footnote{We very grateful to our referee, who noticed that $I_2$ can be reduced to this relatively simple form.} is a bit  more complicated and can be expressed using well known  elliptic integrals:

\bea
I_2 \,\,&=& \,\, \mp \, \frac{\kappa}{2} \int_{u}^{1} \,\sqrt{1 \,\,+\,\,\frac{4 
\,\,\omega\,}{\,\, \kappa^2 \,\,u'\,(1 - u')\,}}\,\, \, du' \,\, \nonumber \\
		\,\,&=& \,\, \mp \, \frac{\kappa}{2} \, \sqrt{\frac{1}{4} + \frac{4 \, \omega}{\kappa^2}} \,\, \left [ \, EE \left ( \, \frac{\pi}{2} \, , \,\, \frac{1}{1 \, + \, 16 \omega / \kappa^2 } \right ) \,\, + \,\, EE \left ( \, ArcSin[1-2u] \, , \,\, \frac{1}{1 \, + \, 16 \omega / \kappa^2 } \right ) \, \right ]
\eea

Where $EE(\varphi, m)$ is  the elliptic  integral of the second   kind \cite{RY}
\beq \label{AN3}
EE(\varphi, m) \,\,\,\equiv\,\,\, \int_{0}^{\varphi} \, \left( 1 - m \cdot \sin^{2} (\theta) \right)^{+ \, 1/2} \, d \theta \;
\eeq

We can see that, in spite the fact that $\phi^{\pm}$ can be calculated explicitly (in radicals), the received expression is to complicated in order to use it for the calculations of the  functions $\Phi^\pm(\omega)$ in \eq{SC5}. 
Nevertheless, this exact calculation very useful by two reasons: (i) using exact form of $\phi^{\pm}$ we can estimate the 
accuracy of our approximate expression for functions $\phi^\pm(\omega,u)$, and 
(ii) we can find the separation parameter between small and large $\omega$ that we used in our semi-classical solution of 
section 3.2.

 The plots in \fig{fig:y2} and in \fig{fig:y1} illustrate how good our approach for small and large $\omega$, which we 
used for calculations of $\phi^{\pm}$.

\FIGURE[ht]{\centerline{\epsfig{file=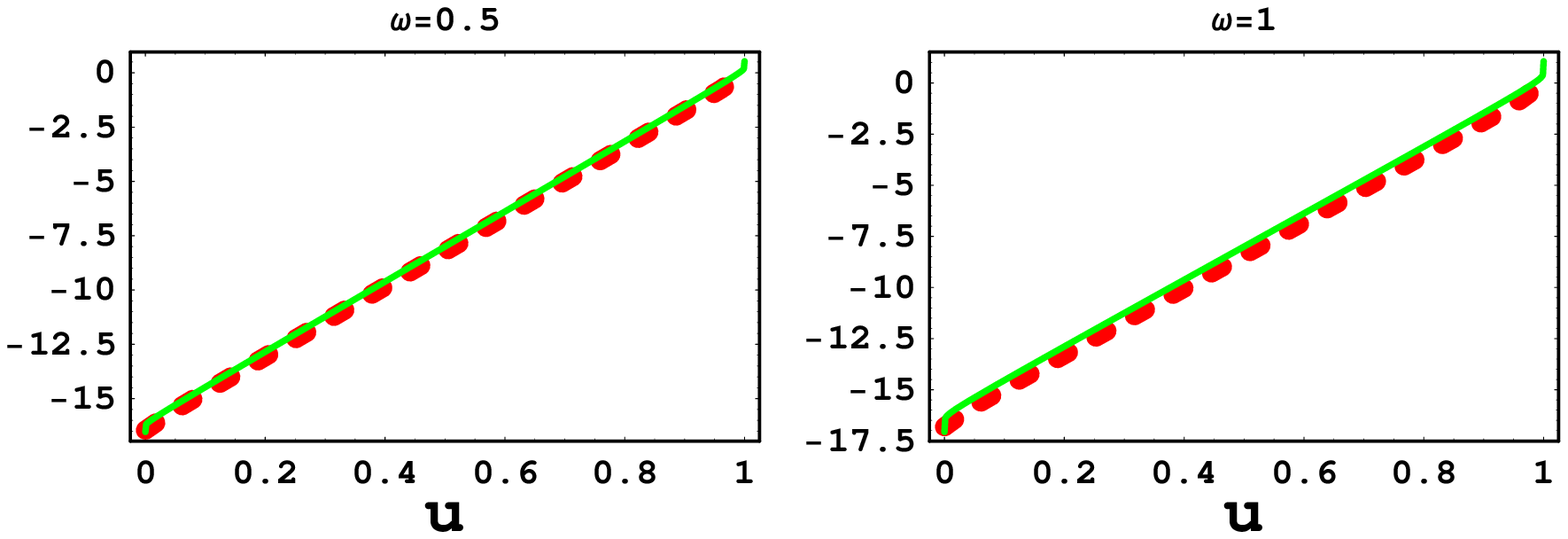,width=15cm, height=6cm}}
\caption{ This plot shows the exact function $ \phi^+(\omega,u)$ 
  (thick dotted line) and  the approximate  function $\phi(\omega)$ of \protect\eq{SC6+} (thin solid line)  for small values of 
$\omega$ (i.e.
 $\omega \,\leq\, \kappa^2$). $\kappa$ = 16   for the plot.}
	\label{fig:y2}}

~

\FIGURE[ht]{\centerline{\epsfig{file=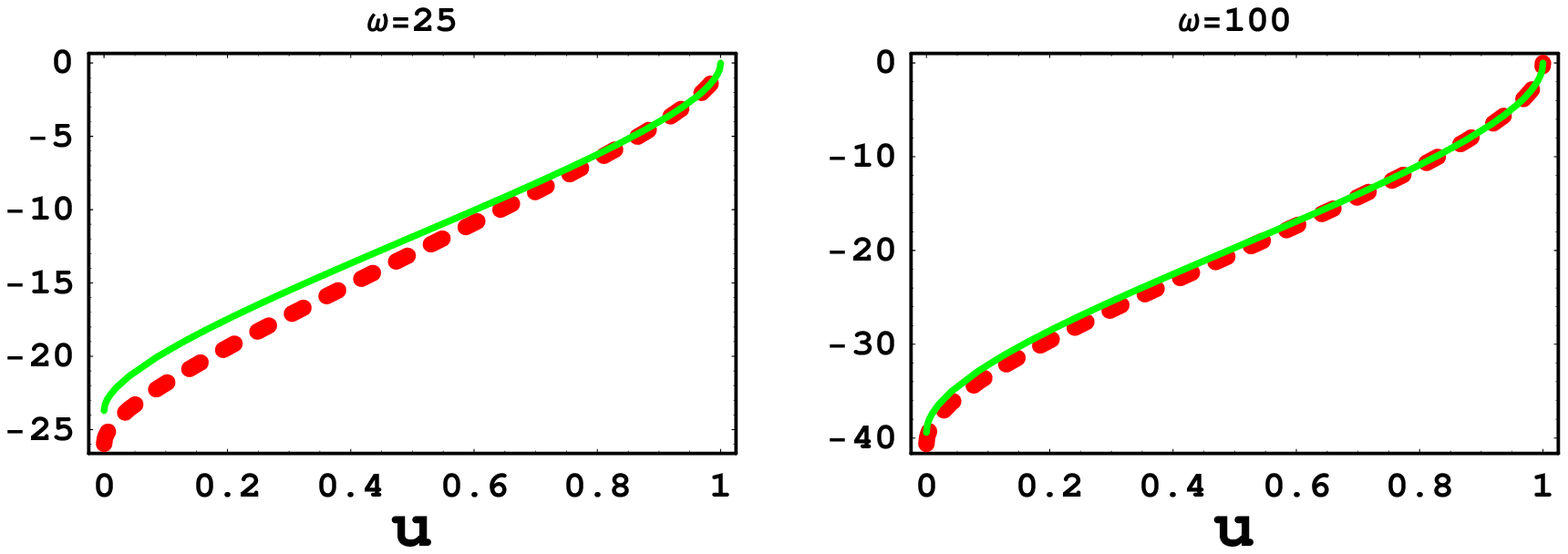,width=15cm, height=6cm}}
\caption{ This plot shows the exact function $ \phi^+(\omega,u)$
  (thick dotted line) and  the approximate  function $\phi(\omega)$ of \protect\eq{SC6+} (thin solid line)  for large values of
$\omega$ (i.e.
 $\omega\, \geq\, \kappa^2$). $\kappa$ = 16 for the plot.}
\label{fig:y1}}

\section{Appendix C: Hermitian Hamultotian description of rapidity evolution.} \label{sec:C}

At this appendix we give the  answer to the   important question: 
whether rapidity evolution in zero transverse dimensions can be described by hermitian Hamiltonian ? Or, in other words: 
does we have probability conservation in these processes ?

\noindent First of all, let us rewrite our master equation \eq{GA1}:

\beq \label{GA1HH}
\frac{\partial Z}{\partial{y}}\,\,=\,\,\,u\,(1 - u)\,\left(-\,\,\kappa\,\,\frac{\partial Z}{\partial u}\,\,+\,\,\frac{\partial^2 Z}{\partial u^2} \right)
\eeq

\noindent in the form of the Fokker-Plank equation, using the simple transformation: $Z(y,u) = u(1-u) \cdot \widetilde{Z}(y,u)$

\beq \label{GA2HH}
\frac{\partial \widetilde{Z}}{\partial{y}}\,\,=\,\,\,-\,\,\,\frac{\partial}{\partial u} \left( \kappa\, u(1-u) \cdot \widetilde{Z}(y,u) \right) \,\,+\,\,\frac{\partial^2 }{\partial u^2}\left(u(1-u) \cdot \widetilde{Z}(y,u) \right) 
\eeq

\noindent It is natural to define the Fokker-Plank operator as 

\beq \label{FP1HH}
\hat{L}_{FP}\, \equiv \,-\,\frac{\partial}{\partial u}a(u) + \frac{\partial^2}{\partial u^2}b(u)
\eeq

\noindent where  in our case $a(u) =  \kappa \, u(1-u)$ and $b(u) = u(1-u)$. Introducing potential $\Phi(u)$ such that:

\beq \label{FPPOT}
\Phi(u) \, \equiv \, \ln [ b(u) ]\, - \, \int_{u_1}^{u} \frac{a(u')}{b(u')} \, du' \, = \, \, \ln [ u(1-u) ] \, - \, \kappa
\eeq

\noindent we can rewrite $\hat{L}_{FP}$ in the following form:

\beq \label{FP2HH}
\hat{L}_{FP} = \frac{\partial}{\partial u}b(u) e^{- \Phi(u)}\frac{\partial}{\partial u}e^{\Phi(u)}
\eeq

\noindent Now it is obviously that $\hat{L}_{FP}$ is non-hermitian operator. Nevertheless, operator $\hat{L}_{H} = e^{\Phi(u) / 
2} \hat{L}_{FP} e^{- \Phi(u) / 2}$ is hermitian (in some sense we restored the hermitian property of our system). Indeed we can 
check this by the direct calculation:

\bea
\int_{u_1}^{u_2} \phi_{1}(u) e^{\Phi(u)} \hat{L}_{FP}  \phi_{2}(u) \, du 
\,\,&=& \int_{u_1}^{u_2} \phi_{1}(u) e^{\Phi(u)} \frac{\partial}{\partial u}b(u) e^{- \Phi(u)}\frac{\partial}{\partial 
u}e^{\Phi(u)}  \phi_{2}(u) \, du  \nonumber \\
&=& - \, \int_{u_1}^{u_2} \left[\frac{\partial}{\partial u}\phi_{1}(u)e^{\Phi(u)}\right]  b(u) e^{- \Phi(u)} \left[\frac{\partial}{\partial u}\phi_{2}(u)e^{\Phi(u)}\right] \, du  \nonumber \\
&=&  \int_{u_1}^{u_2} \phi_{2}(u) e^{\Phi(u)} \frac{\partial}{\partial u}b(u) e^{- \Phi(u)}\frac{\partial}{\partial u}e^{\Phi(u)}  \phi_{2}(u) \, du \, du  \nonumber \\
&=&  \int_{u_1}^{u_2} \phi_{2}(u) e^{\Phi(u)} \hat{L}_{FP}  \phi_{1}(u) \, du  
\eea

\noindent where $\phi_{1}(u)$ and $\phi_{2}(u)$ are two functions, which satisfy boundary conditions. Therefore, we have  
demonstrated 
that $\left( e^{\Phi(u)} \hat{L}_{FP} \right)^{\dagger}  = \hat{L}_{FP}^{\dagger} e^{\Phi(u)} = e^{\Phi(u)} \hat{L}_{FP}$

\noindent It is important to note that transformation $\hat{L}_{H} = e^{\Phi(u) / 2} \hat{L}_{FP} e^{- \Phi(u) / 2}$ can
 be alternatively considered as a transformation of the  appropriate eigenfunctions. If $\phi_{n}(u)$ are the eigenfunctions of 
the 
Fokker-Planck operator $\hat{L}_{FP}$ with the eigenvalues $\lambda_{n}$, the $\psi_{n}(u) = e^{\Phi(u) / 2} \phi_{n}(u)$ will 
be the  
eigenfunctions of operator $\hat{L}_{H}$ with the same eigenvalues. Actually, appropriate transformation of eigenfunctions was 
used in our paper in order to restore Hermitian property of our system. 

\noindent It is obviously that both of system  $\phi_{n}(u)$ and $\psi_{n}(u)$ predefine the orthogonal and complete systems:
\begin{itemize}
	\item Orthogonality: \\ $\int_{u_1}^{u_2} e^{\Phi(u)} \, \phi_{n}(u) \cdot \phi_{m}(u) \, du = \int_{u_1}^{u_2} \psi_{n}(u) \cdot \psi_{m}(u) \, du = \delta_{nm}$ 
	\item Completeness: \\ $e^{\Phi(u)} \sum_{n}  \phi_{n}(u) \cdot  \phi_{n}(u') = \sum_{n}  \psi_{n}(u) \cdot  \psi_{n}(u') = \delta(u - u')$
\end{itemize}

\section{Appendix D: The exact solution as a power series in rapidity.} \label{sec:D}

Our master equation \eq{GA1} can be solved exactly for the arbitrary initial conditions: $Z(Y=0; u) = f(u)$ in terms 
of the  power series of rapidity. This method has a number of exclusive advantages: (I) it  applicable for all 
reasonable 
initial conditions, which allow us to  apply such solution for the different scattering processes at high energies;  (II) with 
this 
method we can easily solve all modifications of our master equation, which incorporates next order corrections
 with the only one restriction that the variable coefficients in the  diffusion equation can be represented as power series in 
$u$); 
(III) 
this method extremely  good for small values of rapidity $y$ and in vicinity of $u=0$ and $u=1$ (we will demonstrate this below 
with numerical calculations). The last argument especially important, since  all solutions, which are based 
on approximate methods, lead to large errors just  for small values of rapidity $y$ and/or in vicinity of $u=0$ and 
$u=1$. 

Unfortunately, solution in terms of power series of rapidity is converged relatively slowly,
 thus this method is not  applicable for calculations of the  asymptotic behaviors of the scattering amplitude. Nevertheless, we 
believe that proposed method will be very useful for practical calculations,
 because most of the experimental data, available now, are  concentrated at small or  intermediate values of rapidity.    

It is very instructive to start with the direct decompositions of function $Z({\cal Y},u)$:

\beq \label{YSER1}
	Z({\cal Y},u) \; = \; \sum_{n=0}^{\infty} C_{n}(u) \cdot {\cal Y}^n
\eeq

\noindent It is obviously the first term in such expansion corresponds
 to appropriate initial conditions, indeed at ${\cal Y} = 
0$ we have:

\beq \label{YSER2}
	Z({\cal Y}=0,u) \; = \; \sum_{n=0}^{\infty} C_{n}(u) \cdot ({\cal Y} = 0)^n \; = \; C_{0}(u) \; = \; f(u)
\eeq

\noindent Now by substitution of \eq{YSER1} into our master equation \eq{GA1} and after arranging the
 appropriate powers of ${\cal Y}$ we will get the following recurrence equations for the functions $C_{n}(u)$:

\beq \label{YSER3}
	C_{n+1}(u) \; = \; \frac{u(1-u)}{n+1} \cdot \left[ \frac{\partial^2 C_{n}(u)}{\partial u^2} - \kappa \frac{\partial 
C_{n}(u)}{\partial u}\right] 
\eeq

\noindent Therefore, once we know initial conditions of our problem \eq{YSER2}, all series expansions
 \eq{YSER1} are well 
defined and can be calculated up to arbitrary order. Finally, introducing the differential operator $\hat L^{n}$ as follows

\beq \label{YSER4}
	\hat L^{i}[f(u)] \; \equiv \; \hat L \left[ \hat L^{i-1}[f(u)] \right] \; ; \;\;\;\;\;\;\;\;\;\;\; \hat L \; \equiv \;
  \frac{u(1-u)}{n+1} \cdot \left[ \frac{\partial^2}{\partial u^2} - \kappa \frac{\partial}{\partial u}\right] 
\eeq

\noindent the answer can be written in the very elegant form: 

\beq \label{YSER5}
	Z({\cal Y},u) \; = \; f(u) \; + \; \sum_{n=1}^{\infty} \hat L^{n} \left[ f(y) \right] \cdot {\cal Y}^n
\eeq

This result is nice from the analytical point of view:
 (1) it haven't singularities, (2) the expansion is well defined for all orders.
 Nevertheless, for numerical estimations probably not so usefully. The reason for this is clear: series
expansion \eq{YSER5} is converged relatively slowly and even more,
 the radius of convergence is the function of $u$. For example, taking first 20 terms of our expansion we can ensure solution 
Z(Y,u) 
in rapidity rang $Y \in [0, 5]$. So from numerical point of view this solution can be considered 
as solution for small and intermediate values of rapidity. 

Fortunately, we can improve numerical property of proposed solution by simple modification, namely we are going expand not 
the function $Z({\cal Y},u)$, but function $\varphi({\cal Y},u) = \ln[Z({\cal Y},u)]$. It is obviously that this modification 
was inspired by our experience in semi-classical solution, but now we are going to keep all terms and propose exact solution.

The first step, which we do is rewriting our master equation \eq{GA1} in terms of 
function $\varphi({\cal Y},u) = \ln[Z({\cal Y},u)]$, by simple substitution we get the following equation for  $\varphi({\cal 
Y},u)$:

\beq \label{YSER6}
\frac{\partial \varphi}{\partial{\cal Y}} \; = \; u (1 - u) \cdot \left( \left( \frac{\partial \varphi}{\partial u} \right)^2 +
  \frac{\partial^2 \varphi}{\partial u^2} - \kappa \frac{\partial \varphi}{\partial u}\right)
\eeq

\noindent Now one again, we search a solution for this equation in terms of power series of rapidity: 

\beq \label{YSER7}
	\varphi({\cal Y},u) \; = \; \sum_{n=0}^{\infty} B_{n}(u) \cdot {\cal Y}^n
\eeq

\noindent   Substituting  \eq{YSER7} into \eq{YSER6} and arranging the appropriate powers of ${\cal Y}$ ,
we  obtain the 
following recurrent relations for functions $B_{n}(u)$:

\beq \label{YSER8}
	B_{n+1}(u) \; = \; \frac{u(1-u)}{n+1} \cdot \left[\sum_{k=0}^{n} \frac{\partial B_{k}(u)}{\partial u}
 \cdot \frac{\partial B_{n-k}(u)}{\partial u} \; + 
\; \frac{\partial^2 B_{n}(u)}{\partial u^2} \; - \; \kappa \frac{\partial B_{n}(u)}{\partial u}\right] 
\eeq

\noindent The appropriate initial condition for function $\varphi({\cal Y},u)$ is also simple $\ln$ transformation:

\beq \label{YSER9}
	\varphi({\cal Y}=0,u) \; = \; \sum_{n=0}^{\infty} B_{n}(u) \cdot ({\cal Y} = 0)^n \; = \; B_{0}(u) \; = \; \ln[f(u)]
\eeq

The last solution numerically is much more stable and, taking first 20 terms of our expansion, we can ensure solution Z(Y,u) 
already in rapidity rang $Y \in [0, 8]$. It is important to note that proposed logarithmic  transformation 
 works  well  in all cases, except the  specific initial conditions: $Z({\cal Y}=0,u)=const$ (fortunately, for our physical 
environment such initial conditions does not look reasonable).

Finally, we demonstrate the quality of this method comparing it with the exact solution of \eq{GS10}.
 It is obviously that for small values of rapidity this method work quite well.
\FIGURE[ht]{
 \centerline{\epsfig{file=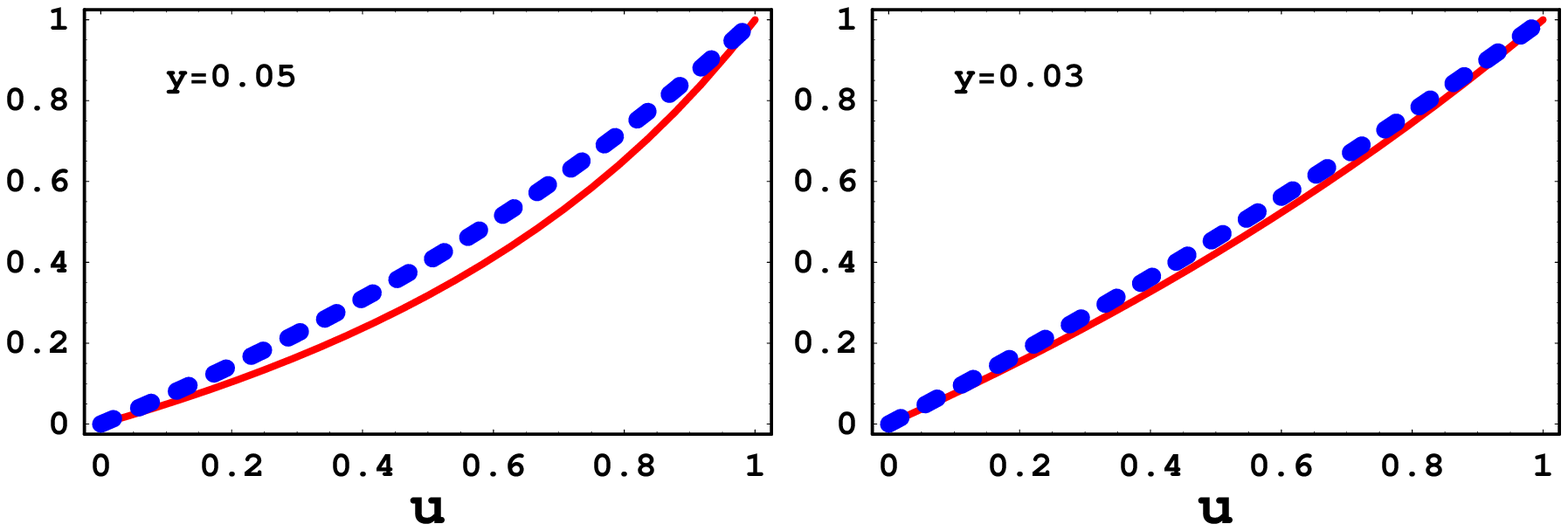,width=16cm, height=6cm}}
\caption{In this plot we compare the exact  solution of \protect\eq{GS10}  - a solid line and the  solution, given by the power 
series 
in rapidity, taking into account first 20 terms (see \protect\eq{YSER5})
 - a dashied line.
 This comparison is performed for two different values of rapidity. As we can see from the plots these solutions are  close.
 In principle, the exact solution is a bit steeper then the  power series solution, which obviously reflects
 the fact that we took into account a  limited number of terms in this expansion.}
 \label{fig:Zcompare}}

\section{Appendix E: The interaction procedure for the spectrum of the problem.}\label{sec:E}

In this appendix we show that the iteration procedure that has been described in 2.3, works quite well at least for 
calculations of the eiegenvalues for the master equation (see \eq{GA1}).  Starting with the exact spectrum of the equation at 
large values of $n$ (see  \eq{GALN1})  and using the set of eigenfunctions we can develop the regular perturbative approach
based on \eq{SCH9} with $U_0\left( \Theta \right) $ being the rectangular-well potential and choosing $V\left( \Theta \right) = 
\frac{\kappa^2}{4}\,\sin^2  \Theta $.  The corrections to the eigenvalues ($\Delta \lambda_n$) are determined by the 
simple formula
\beq \label{C1}
\Delta \lambda_n\,\,\,=\,\,\int^{\pi}_0\,\,d\,\Theta\,\,V\left( \Theta \right)\,|\Psi\left( \Theta \right)|^2
\eeq
and the resulting 
\beq \label{C2}
\lambda_n \,\,=\,\,n^2\,\,+ \,\,\Delta \lambda_n
\eeq
One can see from the Table I that this first iteration leads to the eigenvalues which are very close to the exact eigenvalues 
of the master equation.
\TABLE[ht]{
\begin{tabular}{| l | c | c | c | c | c |c | c | c | c | c |}
\hline \hline
    & & & & & & & & & & \\
Eigenvalues  & $\lambda_1$ & $\lambda_2$ &  $\lambda_3$ & $\lambda_4$ & $\lambda_5$ & $\lambda_6$ & $\lambda_7$ & $\lambda_8$ 
& $\lambda_9$ & $\lambda_{10}$\\
\hline
    & & & & & & & & & & \\
 Spheroidal (exact) & 13.0 & 14.13 & 21.40 & 28.63 & 38.45 & 50.32 & 64.24 & 80.19 & 98.15 & 118.13\\
\hline
    & & & & & & & & & & \\
Asymptotic ($\lambda_n = n^2$) & 1 & 4 & 9 & 16 & 25  & 36 & 49 & 64 & 81 & 100 \\
\hline
    & & & & & & & & & & \\
First iteration & 14.25 & 14.75 & 21.25 & 29.75 & 40.25 & 52.75 & 67.25 & 83.75 & 102.25 & 122.75\\
 & & & & & & & & & & \\
\hline \hline
\end{tabular}
\caption{The comparison of the first ten eigenvalues of the master equation with the asymptotic estimates and with the first 
iteraction described in section 2.3.}
}

From this table we see that the first iteration reproduces correctly the values of the eigenvalues at small $n$. Therefore, we 
can conclude that the iteration procedure, described in section 2.3, gives the way to approach the exact solution with a good 
accuracy.

\end{document}